\documentclass[12pt]{article}
\usepackage{amsmath}
\usepackage{graphicx,psfrag,epsf}
\usepackage{enumerate}
\usepackage{url} 
\usepackage[utf8]{inputenc}
\usepackage[backend=biber, style=authoryear, natbib=true, mincitenames=1, maxcitenames=1, maxbibnames=99]{biblatex}

\usepackage[T1]{fontenc}
\usepackage{lmodern}
\usepackage{caption}
\usepackage{csquotes}
\usepackage[english]{babel}
\usepackage{afterpage}
\usepackage{abstract}
\usepackage[obeyDraft, bordercolor=red, backgroundcolor=white]{todonotes}
\usepackage{amssymb, amsfonts, amsmath, amsthm}              
\usepackage{graphicx} 
\usepackage{subcaption}
\usepackage[toc,page]{appendix}
\usepackage[hidelinks]{hyperref}
\usepackage{bm}
\usepackage{colortbl}
\usepackage{longtable}
\usepackage{rotating}
\usepackage{lscape}
\usepackage{setspace}
\usepackage{flafter}
\usepackage{morefloats}
\usepackage{array}
\usepackage{tabularx,booktabs}
\usepackage{authblk}
\usepackage{xr}
\usepackage{mathtools}
\usepackage{placeins}
\usepackage{enumitem}
\usepackage[normalem]{ulem}
            
\newcommand\citeS[1]{\citeauthor{#1}'s\ (\citeyear{#1})}

\newtheorem{assumption}{Assumption}
\newtheorem{corollary}{Corollary}
\newtheorem{theorem}{Theorem}
\newtheorem{property}{Property}
\newtheorem{proposition}{Proposition}
\newtheorem{lemma}{Lemma}
\newtheorem{remark}{Remark}
\newtheorem{algorithm}{Algorithm}
\DeclareMathOperator*{\argmin}{argmin}
\newtheorem{definition}{Definition}

\newtheorem{intassumption}{Assumption}
\numberwithin{intassumption}{assumption}
\newcounter{referredassumption}

\newcommand{\blind}{0}

\newcolumntype{C}[1]{>{\centering\let\newline\\\arraybackslash\hspace{0pt}}m{#1}}

\begin{document}

\def\spacingset#1{\renewcommand{\baselinestretch}%
{#1}\small\normalsize} \spacingset{1}


\if0\blind
{
  \title{\bf Agglomerative Hierarchical Clustering\\ for Selecting Valid Instrumental Variables}
  \author[1]{Nicolas Apfel}
  \author[2]{Xiaoran Liang\thanks{Corresponding address: n.apfel@soton.ac.uk, University of Southampton - Department of Economics
  		Murray Building (B58), University Road, Southampton, SO17 1BJ, United Kingdom. Nicolas Apfel gratefully acknowledges funding funding through the International Doctoral Program ``Evidence-Based Economics'' of the Elite Network of Bavaria and through ESRC grant EST013567/1. Xiaoran Liang acknowledges support from the Medical Research Council grant MC/MR/WO14548/1, the ESRC grant ES/P000630/1, and from the Jean Golding Institute PGR Seed Corn Funding. The authors would like to thank the editor, three anonymous referees, Frank Windmeijer, as well as seminar participants at various conferences and seminars for providing helpful comments.}}
  \affil[1]{Department of Economics, University of Southampton, UK} 
  \affil[2]{Exeter Medical School, University of Exeter, UK}
  \setcounter{Maxaffil}{0}
  \renewcommand\Affilfont{\itshape\small}
  \maketitle
} \fi

\if1\blind
{
  \bigskip
  \bigskip
  \bigskip
  \begin{center}
    {\LARGE\bf Title}
\end{center}
  \medskip
} \fi

\bigskip
\begin{abstract}
\noindent We propose a procedure which combines hierarchical clustering with a test of overidentifying restrictions for selecting valid instrumental variables (IV) from a large set of IVs. Some of these IVs may be invalid in that they fail the exclusion restriction. We show that if the largest group of IVs is valid, our method achieves oracle properties. Unlike existing techniques, our work deals with multiple endogenous regressors. Simulation results suggest an advantageous performance of the method in various settings. The method is applied to estimating the effect of immigration on wages. \medskip 
\end{abstract}

\noindent%
\vfill

\newpage

\spacingset{1.8} 
\section{Introduction}\label{sec:intro}
 
Instrumental variables estimation is a widely used method for analysing the causal effects of treatment variables on an outcome when the causal relationship between them is confounded. Consistent estimation requires all instruments to be valid. This requires that

\begin{enumerate}[label=(\alph*)]
    \item Relevance: Instruments are associated with the endogenous variables.
    \item Exclusion: IVs do not affect the outcome directly or through unobserved factors.
\end{enumerate}
In this paper, we propose a new method to select the valid instruments, by combining the agglomerative hierarchical clustering (AHC) algorithm, a statistical learning method commonly employed in cluster analysis, with the Hansen-Sargan test \citep{Sargan1958Estimation, Hansen1982Large} of overidentifying restrictions. We rely on the plurality rule \citep{Guo2018Confidence} which states that the largest group of IVs is valid.  Instruments are said to form a group if their IV-specific just-identified estimators converge to the same value. Under plurality, our method achieves oracle selection, meaning that valid instruments can be selected consistently, and the two-stage least squares (2SLS) estimator using the instruments selected as valid has the same limiting distribution as the ideal estimator that uses the set of truly valid instruments. 

Previous work has tackled the IV selection problem in the single endogenous variable case.  \citet{Kang2016Instrumental} propose a selection method based on the least absolute shrinkage and selection operator (LASSO). \citet{Windmeijer2019Use} make improvements by proposing an adaptive Lasso based method that has oracle properties under the assumption that more than half of the candidate instruments are valid (the \textit{majority} rule). \citet{Guo2018Confidence} propose the Hard Thresholding with Voting method (HT) that has oracle properties under the sufficient and necessary identification condition that the largest group is formed by all the valid instruments (the \textit{plurality} rule). This is a relaxation to the majority rule. Under the same identification condition,  \citet{Windmeijer2021Confidence} propose the Confidence Interval method (CIM), which has better finite sample performance. 

\noindent
Our research adds to the literature in two ways:
\begin{enumerate}
	\item  We combine agglomerative hierarchical clustering with a traditional statistical test, the Sargan overidentification test, to yield a novel algorithm for IV selection. 
	\item We extend our method to accommodate multiple endogenous regressors. This is not available for the aforementioned methods, but it is straightforward in our setting. 
\end{enumerate}
The new method provides the theoretical guarantee that under the plurality rule it can select the true set of valid instruments consistently, which is also true for HT and CIM for the one regressor case. 
In Monte Carlo simulations, we show that our method achieves oracle properties in the multiple endogenous regressors setting (Section \ref{sec:MonteCarlo}). We illustrate the consequences of local-to-zero violations of exclusion and weak IVs. In the Appendix, we compare our method with HT and CIM in the single regressor case, which also rely on the plurality rule and show that our method shows comparable performance with strong IVs, and has an edge on them in the case with weak IVs. It is worth mentioning that AHC is computationally less complex than CIM and HT.

Our work adds to a growing literature on valid IV selection inspired by \citet{Andrews1999Consistent} who proposes moment selection criteria and a downward testing procedure. In our context, the number of IVs is large in the sense that it  exceeds the number of regressors and considering all possible overidentified models is infeasible. However, we are not in a setting where the number of IVs grows with the sample size. Our setting is also different from the one in \citet{Belloni2012Sparse} that uses regularization to find an optimal set of instruments, in that it does not uphold the assumption that all IVs fulfil the exclusion restriction. 

We illustrate the various strengths of our method by revisiting the estimation of the effect of immigration on wages in the US. The results indicate that the actual effects might be much larger than suggested by standard 2SLS estimates. 
Researchers often use previous shares of immigrants as IVs and try to estimate the effects of current and past migration contemporaneously, making this a good illustration for the case with multiple regressors. 
We also provide R-code that makes implementation easy in practice.

The remainder of this paper is structured as follows. Section \ref{sec:ModelAndAssumptions} states the model and assumptions, illustrates some well-established properties of the just-identified estimator and introduces the main theoretical properties. Section \ref{sec:IVSelectionAndEstimation} describes the method and the algorithm when there are multiple endogenous variables, and investigates its asymptotic properties. In Section \ref{sec:MonteCarlo}, we provide Monte Carlo simulation results. Section \ref{sec:migration} revisits the effect of immigration on wages in the US. Section \ref{sec:Conclusion} concludes. The appendix in the supplementary material includes discussions, all proofs and additional simulations. 

\section{Model and Assumptions}\label{sec:ModelAndAssumptions}

In the following, we introduce notational conventions used throughout this paper. Matrices are in upper case and bold, vectors are in lower case and bold, scalars are in lower case and not in bold. 
Let $\mathbf{y}$ be an $n \times 1$-vector of the observed outcome, 
$\mathbf{d}_1$, ..., $\mathbf{d}_P$ be $P$ endogenous regressor vectors (each $n \times 1$), which can be subsumed in an $n \times P$ - matrix $\mathbf{D}$, and $\mathbf{z}_1$, ..., $\mathbf{z}_J$ be $J$ instrument vectors, which can be subsumed in an $n \times J$ - matrix $\mathbf{Z}$.
Let error terms be $\mathbf{u}$ and $\bm{\varepsilon_p}$ for $p \in \{1, ..., P\}$, which are all $n \times 1$ error-vectors and are correlated with $\sigma_{up} := cov(\mathbf{u}, \bm{\varepsilon}_p)$. The latter covariances measure the endogeneity of regressors in $\mathbf{D}$. The $P \times 1$ coefficient vector of interest is $\bm{\beta}$. 
The $J \times P$ matrix $\bm{\gamma}$ contains first-stage coefficients.\footnote{To be consistent with the literature, we denoted this matrix as lower case because upper case $\mathbf{\Gamma}$ denotes the reduced form parameters.}
Let $s$ be the number of instruments in the set of valid instruments, $\mathcal{V}$, $g$ be the number of instruments in the set of invalid instruments, $\mathcal{I}$, and $J =  g + s$ be the total number of instruments in the overall set of instruments, $\mathcal{J}$. 
The arithmetic mean of a variable $x$ is defined as $\mu_x = \frac{\Sigma_{i=1}^{n} x_i}{n}$, the mean of a vector is the vector of dimension-wise arithmetic means, $\lVert \cdot \rVert$ denotes the L2-norm and $|\cdot|$ denotes cardinality, when used around a set and an absolute value, when used around a scalar. The symbol $\wedge$ denotes the logical conjunction, \textit{and}. The $n \times n$ projection matrix is $\mathbf{P}_X = \mathbf{X}(\mathbf{X}'\mathbf{X})^{-1}\mathbf{X}'$, and the annihilator matrix is $\mathbf{M}_X = \mathbf{I} - \mathbf{P}_X$ and $\hat{\mathbf{D}} = \mathbf{P}_Z \mathbf{D}$ are the fitted values. Throughout the paper, we assume that $J$ and $g$ are fixed and $P < J$. 

\subsection{Model Setup}\label{sec:ModelSetup}

The following observed data model takes the potentially invalid instruments into account:
\begin{equation}\label{eq:Structural}
\mathbf{y} = \mathbf{D} \bm{\beta} + \mathbf{Z} \bm{\alpha} + \mathbf{u} \text{,}
\end{equation}
with $\mathbf{E}[u_i | \mathbf{z}_i] = 0$. This observed data model is the same as in \citet{Kang2016Instrumental, Guo2018Confidence, Windmeijer2021Confidence} who derive it from a potential outcomes model. The linear projection of $\mathbf{D}$ on $\mathbf{Z}$ is
\begin{equation}\label{eq:FirstStage}
\mathbf{D} = \mathbf{Z} \bm{\gamma} + \bm{\varepsilon}.
\end{equation}

\noindent The vector $\bm{\alpha}$ is $J \times 1$ with entries $\alpha_j$, each associated with an individual IV. Each entry indicates which of the IVs has a direct effect on the outcome variable and hence is invalid. Following a large econometric and statistical literature, such as \citet{Conley2012Plausibly}, \citet{Kang2016Instrumental}, \citet{Guo2018Confidence} or \citet*{Masten2021Salvaging} we define a valid IV as: 
\begin{definition}
For $j = 1, ..., J$, $\mathbf{z}_j$ is valid if $\alpha_j = 0$. If $\alpha_j \neq 0$, then $\mathbf{z}_j$ is invalid.
\end{definition}
\noindent Following the literature, we restrict our attention to violations of the exclusion restriction. This could be extended to exogeneity violations, as in $Cov(\mathbf{Z},\mathbf{u})\neq0$. The consequences of different violations are beyond the scope of this work and have been discussed in \citet{Apfel2022}. 

The ideal model which selects the truly valid instruments as valid and controls for the set of invalid instruments is the oracle model, defined as 
\begin{equation}\label{eq:Oracle}
\mathbf{y} = \mathbf{D} \bm{\beta} + \mathbf{Z}_{\mathcal{I}} \bm{\alpha}_{\mathcal{I}} + \mathbf{u} = \mathbf{X}_{\mathcal{I}} \bm{\theta}_{\mathcal{I}} + \mathbf{u} \text{,}
\end{equation}
where $\mathbf{X}_\mathcal{I} = (\mathbf{D} \quad \mathbf{Z}_\mathcal{I})$ and $\bm{\theta}_\mathcal{I} = (\bm{\beta} \quad \bm{\alpha}_\mathcal{I}')'$.

\subsection{Assumptions}\label{sec:Assumptions}

The assumptions that follow are analogous to the ones in \citet{Windmeijer2021Confidence}, but for the general case $P\geq1$. The inputs of our method, all the just-identified estimators, are estimated by all the $P$-combinations from $\mathbf{z}_1, ..., \mathbf{z}_J$. Hence we have $\binom{J}{P}$ just-identified estimators. This simplifies to $J$ estimators when $P=1$. Let $[j]$ be a set of identities of any $P$ instruments such that the model is exactly identified with these $P$ instruments. Let $\mathbf{Z}_{[j]}$ denote the corresponding $n \times P$ instrument matrix. 
\begin{assumption}\label{ass:RankAssumption}
	Rank assumption.
	$$E(\mathbf{z}_i \mathbf{z}_i') = \mathbf{Q} \text{ with } \mathbf{Q} \text{ a finite and full rank matrix.}$$
\end{assumption}
\begin{assumption}\label{ass:FirstStageSingle} Identification of just-identified models.\\
		For all possible $[j]$, let $\bm{\gamma}_{[j]}$ be the combination of the $k_{th}$ rows of $\bm{\gamma}$ for $k \in [j]$. We assume $$rank(\bm{\gamma}_{[j]})=P.$$
\end{assumption}
\begin{assumption}\label{ass:ErrorStructure} Error structure.\\
	Let $\mathbf{w}_i = (u_i \quad \varepsilon_{i,p})'$ for $p=\{1,...,P\}$. Then, $E(\mathbf{w}_i)=\mathbf{0}$ and\\ $E[\mathbf{w}_i \mathbf{w}_i']= \left( \begin{array}{rr}
		\sigma_u^2 & \sigma_{u, \varepsilon_p} \\
		\sigma_{u, \varepsilon_p} & \sigma_{\varepsilon_p}^2 
	\end{array}\right)
	= \bm{\Sigma}_p $ with 
	$Var(u_i) = \sigma_u^2, \,\, Var(\varepsilon_i) = \sigma_{\varepsilon}^2$ $Cov(u_i, \varepsilon_{i,p}) = \sigma_{u, \varepsilon_p}$ and the elements of $\bm{\Sigma}$ are finite.
\end{assumption}
\begin{assumption}\label{ass:Asymptotics}
	\begin{align}
	plim (n^{-1} \mathbf{Z}' \mathbf{Z}) = E(\mathbf{z}_i \mathbf{z}_i') = \mathbf{Q}_{ZZ}
	\quad &; \quad plim(n^{-1} \mathbf{Z}' \mathbf{D}) = E(\mathbf{z}_i \mathbf{d}_{i}') = \mathbf{Q}_{ZD} \notag\\
	plim(n^{-1} \mathbf{Z}' \bm{u})= E(\mathbf{z}_i u_i) = 0 \quad &;  \quad
	plim(n^{-1} \mathbf{Z}' \bm{\varepsilon}) = E(\mathbf{z}_i \bm{\varepsilon}_i) = \mathbf{0} \notag\\
	plim(n^{-1}\sum\limits_{i=1}^{n} \mathbf{w}_i) = 0 \quad &;  \quad plim(n^{-1} \mathbf{w}_i \mathbf{w}_i') = \bm{\Sigma} \text{.}  \notag
	\end{align}
\end{assumption}
\begin{assumption}\label{ass:ZwNormal}
$\frac{1}{\sqrt{n}}\sum\limits_{i=1}^n vec(\mathbf{z}_i \mathbf{w}_i') \overset{d}{\rightarrow} N(0,\bm{\Sigma \otimes \mathbf{Q}}) \text{ as } n\rightarrow \infty$.
\end{assumption}
\noindent Assumptions \ref{ass:RankAssumption} and \ref{ass:FirstStageSingle} guarantee identification for all $\binom{J}{P}$ models. Assumption \ref{ass:ZwNormal} is made for ease of exposition, but the method can be easily extended to accommodate heteroskedasticity, clustering and serial correlation. 
From (\ref{eq:Structural}) and (\ref{eq:FirstStage}), we have the outcome-instrument reduced form
\begin{equation*}
\mathbf{y} = \mathbf{Z}\bm{\Gamma} +\bm{\epsilon}
\end{equation*}
where $\bm{\Gamma} = \bm{\gamma} \bm{\beta} + \bm{\alpha}$.
Each just-identifying combination of IVs ${[j]}$ is associated with a just-identified estimator $\hat{\bm{\beta}}_{[j]}$, the 2SLS estimator using $\mathbf{Z}_{[j]}$ as the just-identifying combination of IVs, and controlling for the rest. We write the $\binom{J}{P}$ just-identified estimators as:
\begin{equation*}
\hat{\bm{\beta}}_{[j]}= \hat{\bm{\gamma}}_{[j]}^{-1}\hat{\bm{\Gamma}}_{[j]} 
\end{equation*}
where $\hat{\bm{\Gamma}}_{[j]}$ and $\hat{\bm{\gamma}}_{[j]}$ are the OLS estimators for $\bm{\Gamma}_{[j]}$ and $\bm{\gamma}_{[j]}$, i.e. the ${[j]}$-th entries and rows of $\bm{\Gamma}$ and $\bm{\gamma}$ respectively. Details can be found in Appendix \ref{app:PropertiesOfJustIdentified}. Then we have

\begin{property}Properties of just-identified estimators.\label{prop:JustIdentified}\\
Under Assumptions \ref{ass:FirstStageSingle} to \ref{ass:ZwNormal} it holds that
	$$plim\bm{ \hat{\beta}}_{[j]} = \bm{\beta} + \bm{\gamma}_{[j]}^{-1}\bm{\alpha}_{[j]} = \bm{\beta} + \mathbf{q}_{[j]}$$
\end{property}
\noindent where the inconsistency is $plim\hat{\bm{\beta}}_{[j]} - \bm{\beta} = \bm{\gamma}_{[j]}^{-1}\bm{\alpha}_{[j]} = \mathbf{q}_{[j]}$ and there are $\binom{J}{P}$ inconsistency terms $\mathbf{q}_{[j]}$. In the single regressor case, this becomes $plim(\hat{\beta}_{j})= plim\left(\frac{\hat{\Gamma}_j}{\hat{\gamma}_j}\right) = \beta + \frac{\alpha_j}{\gamma_j}$ with inconsistency $plim(\hat{\beta}_{j}) - \beta = \frac{\alpha_j}{\gamma_j} = q_j$. For $P=1$, \citet{Guo2018Confidence} define a group as:

\begin{definition}\label{def:Group}
A group $\mathcal{G}_q$ is a set of IVs that has the same estimand $\beta_j = \beta + q$.
$$\mathcal{G}_{q} = \{j: \beta_j = \beta + q\}$$
\end{definition}
\noindent where $q \in \mathbb{R}$.
In words, a group is a set of IVs whose just-identified estimators converge to the same value $\beta + q$, where $q$ loses the subscript because it is the same for all IVs in the group. The group consisting of all valid IVs is $\mathcal{G}_0 = \{j: q_j=0\}$. Let the number of groups be $G$. 
The plurality assumption in \citet{Guo2018Confidence} is key and it states that among the $G$ groups formed by $\mathbf{z}_1, ..., \mathbf{z}_J$, the largest one is composed by all valid IVs. 
\begin{assumption}\label{ass:plurality}Plurality Rule.\\
	$$g > \underset{q\neq0}{max} |\mathcal{G}_q| $$
\end{assumption}
\noindent where $g$ is the number of valid instruments. 
Because when $P>1$, each IV is not associated with a single scalar $q$, we introduce the concept of a \textit{family}: 
\begin{definition}\label{def:family}
	A family is a set of just-identifying IV combinations that is associated with just-identified estimators which converge to the same value.
	$$\mathcal{F}_{q} = \{[j]: \bm{\beta}_{[j]}=\bm{\beta} + \mathbf{q}\}.$$
\end{definition}
\noindent Let there be $Q$ families. The family that consists of IV combinations which generate consistent estimators is $\mathcal{F}_0 = \{[j]: \mathbf{q}=\mathbf{0}\}$. A natural extension of the plurality assumption to the family case is $|\mathcal{F}_0|  > \underset{q \neq 0}{max} |\mathcal{F}_q|$. We show in Appendix \ref{app:gP}, that a combination of IVs is an element of $\mathcal{F}_0$ if and only if all of the IVs in the combination are valid. This means that the family with $\mathbf{q}=\mathbf{0}$ consists of all combinations that use IVs from the set of valid instruments, $\mathcal{V}$, and hence $|\mathcal{F}_0| = \binom{g}{P}$. Therefore, the plurality assumption becomes

\setcounter{referredassumption}{6}
\setcounter{intassumption}{0}
\begin{intassumption}\label{ass:familyplurality} Family plurality\\
	$$ \binom{g}{P} > \underset{\mathbf{q}\neq\mathbf{0}}{max} |\mathcal{F}_q|.$$
\end{intassumption}
\noindent 
The inconsistency term of elements in $\mathcal{F}_{q}$ with $\mathbf{q} \neq \mathbf{0}$ depends on both $\bm{\gamma}_{[j]}$ and $\bm{\alpha}_{[j]}$ and hence there is no direct relation from $\bm{\alpha}_{[j]}$ to $\mathbf{q}_{[j]}$, unless $\bm{\alpha}_{[j]}=\mathbf{0}$. 
One way in which this new plurality rule can be fulfilled, is when the largest set of IVs has zero direct effects $\alpha_j=0$ and the vectors $\bm{\gamma_{[j]}^{-1}} \bm{\alpha_{[j]}}$ constituted by $P$-sets with $\bm{\alpha_{[j]}}\neq 0$ are sufficiently dispersed. Strictly speaking, the family plurality assumption can also hold when the largest set of IVs has some non-zero effect $\alpha_j$. If the dispersion of $\bm{\gamma_{[j]}^{-1}} \bm{\alpha_{[j]}}$ is large enough, the largest family will still be constituted by valid IVs only. 

The plurality assumption is reasonable when researchers can credibly uphold validity, but might have missed some direct effect between some IV and the outcome because perfect structural knowledge of the mechanisms is not given. In our application, all instruments follow a similar logic but some are long lags, while others are short lags of a variable. If the violation of validity comes from correlation with unobservables and these are serially correlated, it might make sense to believe that serial correlation breaks with long lags. More detailed discussion of the plurality assumption can be found in section \ref{sec:migration}.
\section{IV Selection and Estimation Method}\label{sec:IVSelectionAndEstimation}

Based on the definition of groups, families and the plurality rule, a natural strategy for IV selection is to find out the $G$ IV groups and then select the largest group as the set of valid instruments. 
The CIM and HT also build on the plurality assumption, but they are presented in the case $P=1$, which we would like to extend. The CIM is based on ordered, overlapping confidence intervals, but in the multiple regressor case it is not clear how to order and find overlapping confidence regions. The HT is based on pairwise tests for the Null hypotheses $\beta_j=\beta_k$. The method could in fact be extended, but there are a number of disadvantages: first, one would need to run $\binom{J}{P}\cdot\left(\binom{J}{P}-1\right)$ pairwise tests, while AHC needs at most $\binom{J}{P}$ tests. Second, HT does not include a downward testing procedure and it is not clear how to incorporate it to robustly select the tuning parameter. Third, HT can lead to counter-intuitive selection results: HT selects a set of IVs which has been voted for by other IVs, but in finite samples the IVs involved in the voting do not have to be the ones selected as valid. This can lead to situations where pairwise tests of the estimators in the largest group in fact do not agree with each other in that a pairwise test would lead to rejection. In AHC, all point estimates need to be close to each other by construction. Finally, the link between largest groups of point estimates and instrument groups would not be immediate. We have provided such a link with our family plurality assumption and our selection method and these concepts can also be helpful when trying to extend HT to the multiple regressor case.
 
In this paper, we explore the clustering methods to discover the IV groups. First, we fit the general clustering framework to the IV selection problem, which is summarized in the minimisation problem in (\ref{eq:min}). This general method needs a pre-specified parameter $K$, which is the number of clusters. We show that when $K$ equals the true number of families, $K=Q$, asymptotically there is a unique solution to this minimization problem, and this solution coincides with the true underlying partition. However, the fact that consistent selection depends on $K$ makes it difficult to implement in practice, as we do not have prior knowledge about the number of families. If $K$ is too large (larger than the number of families), then the largest family will be split. If $K$ is too small, then the largest group might be in a cluster with some other family. To tackle this problem, we propose a downward testing procedure which combines agglomerative hierarchical clustering (Ward's method) with the Sargan test for overidentifying restrictions to select the valid instruments, which allows us to select the valid instruments without pre-specifying $K$.

\subsection{Clustering Method for IV Selection}\label{sec:ClusteringMethod}

Let $\mathbf{\mathcal{S}}(K) = \{\mathcal{S}_1, ..., \mathcal{S}_K\}$ be a partition of $\binom{J}{P}$ just-identified estimators $\hat{\bm{\beta}}_{[j]}$ into $K$ cluster cells with identities $k = 1,..., K$. This results in the following minimization problem:
\begin{equation}\label{eq:min}
\hat{\mathbf{\mathcal{S}}}(K)= \underset{\mathbf{\mathcal{S}}}{\argmin} \sum_{k=1}^{K}\sum_{\hat{\bm{\beta}}_{[j]} \in \mathcal{S}_k}||\hat{\bm{\beta}}_{[j]} - \bar{\mathcal{S}}_k||^2 \text{,}
\end{equation}
where $\bar{\mathcal{S}}_k$ is the arithmetic mean of all just-identified estimators in cluster $\mathcal{S}_k$. The term $\sum_{k=1}^{K}\sum_{\hat{\bm{\beta}}_{[j]} \in \mathcal{S}_k}||\hat{\bm{\beta}}_{[j]} - \bar{\mathcal{S}}_k||^2$ in equation \ref{eq:min} corresponds to the intra-cluster variance, summed over the number of clusters. Based on Assumption \ref{ass:familyplurality}, for a given $K$, the cluster that consists of just-identified estimators obtained with valid IVs is estimated as the cluster that contains the largest number of estimators: 
$$\hat{\mathcal{S}}_m(K) = \{ \hat{\mathcal{S}}_k(K): |\hat{\mathcal{S}}_k(K)| = \underset{k}{max}|\hat{\mathcal{S}}_k(K)|\}.$$
The elements in $\hat{\mathcal{S}}_m(K)$ are just-identified estimates which need to be translated to \textit{families}, $\hat{\mathcal{F}}(K)$, which are sets of IV combinations. The family associated with the largest cluster is
$$\hat{\mathcal{F}}_m(K) = \{[j]: \hat{\bm{\beta}}_{[j]} \in \hat{\mathcal{S}}_m(K)\}.$$
Now, the families need to be translated to sets of IVs to be tested. To achieve this, for each $K$, the potentially valid IVs are selected as those that are in the largest family.
$$\hat{\mathcal{V}}_m(K) = \{j:[j] \in \hat{\mathcal{F}}_m(K)\}.$$
The remaining IVs are then selected as invalid:
$$\hat{\mathcal{I}}(K) = \mathcal{J}\quad \backslash \quad \hat{\mathcal{V}}_m(K).$$
\noindent 
For $P=1$, each cluster is directly associated with a group of IVs, and we can select $\mathcal{V}$ as 
$\hat{\mathcal{V}}(K) = \{j: \hat{\beta}_j \in \hat{\mathcal{S}}_m(K)\} \text{.}$
For each $K$, in the finite sample there might be cases where there are multiple maximal clusters $\hat{\mathcal{S}}_m(K)$ and multiple $\hat{\mathcal{V}}_m(K)$. Let $\hat{\mathcal{V}}^M(K)$ denote the set of the multiple $\hat{\mathcal{V}}_m(K)$. 
In this case, we select the cluster in which the most IVs are involved, $\hat{\mathcal{V}}^{Max}(K) = \{\hat{\mathcal{V}}_m(K): |\hat{\mathcal{V}}_m(K)| = \underset{}{max} |\hat{\mathcal{V}}^M(K)|\}$. If there are multiple clusters with maximal number of estimates \textit{and} IVs, we select the set of IVs which leads to the lowest Sargan statistic. Then for each $K$, the unique set to be tested is: 
\begin{equation}\label{eq:SelectedSargan}
	\hat{\mathcal{V}}^{Sar}(K) = \{\hat{\mathcal{V}}_m(K):  Sar(\hat{\mathcal{V}}_m(K))=min \,Sar(\hat{\mathcal{V}}^{Max}(K))\}.
\end{equation}
When the number of clusters $K$ is equal to the number of families $Q$, $K=Q$, then for $n\rightarrow\infty$ there is a partition minimizing the sum in Equation (\ref{eq:min}). Asymptotically, this occurs, when the grouping is such that $\hat{\mathcal{S}}_k = \mathcal{S}_q$ and each selected family $\hat{\mathcal{F}}_k$ is in fact formed by a true family, $\mathcal{F}_q$. Define the partition leading to this grouping as the true partition $\mathbf{\mathcal{S}}_0 = \{\mathcal{S}_{01}, ..., \mathcal{S}_{0Q} \}$. 
To see that, first note that if the partition is such that $\hat{\mathcal{S}}_k = \mathcal{S}_{0q} \,\, \forall k, q$, i.e. $\hat{\mathbf{\mathcal{S}}}(K) = \mathbf{\mathcal{S}}_0$,
\begin{equation*}
g(\hat{\mathbf{\mathcal{S}}}(K)) = g(\mathbf{\mathcal{S}}_0) = plim \{\sum_{k=1}^{K}\sum_{\hat{\bm{\beta}}_{[j]} \in \mathcal{S}_k}||\hat{\bm{\beta}}_{[j]} - \bar{\mathcal{S}}_k||^2\} = 0 \text{,}
\end{equation*}
where $g(\cdot)$ is the $plim$ of the sum of squared deviations from the cluster mean, summed over $k$. 
For all $\hat{\bm{\beta}}_{[j]} \in \mathcal{S}_{0k}$, we have $plim \, \hat{\bm{\beta}}_{[j]} = plim \, \bar{\mathcal{S}}_k $, and $plim\{||\hat{\bm{\beta}}_{[j]} - \bar{\mathcal{S}}_k||^2\} = 0$. This is the case for all $k \in 1,...K$, hence $g(\mathbf{\mathcal{S}}_0) = 0$. If the partition is such that some $\mathcal{S}_k \neq \mathcal{S}_{0q}$, i.e. $\mathbf{\mathcal{S}} \neq \mathbf{\mathcal{S}}_0$, then $plim \, \hat{\bm{\beta}}_{[j]} \neq plim \, \bar{\mathcal{S}}_k $ for some $\hat{\bm{\beta}}_{[j]} \in \mathcal{S}_k$ and $g(\mathbf{\mathcal{S}}) > 0$. This means that when $n \rightarrow \infty$ and $K=Q$ there is a unique solution for Equation \ref{eq:min}, which is such that $\mathbf{\mathcal{S}}=\mathbf{\mathcal{S}}_0$. 

\subsection{Ward's Algorithm for IV Selection}

To choose the correct value of $K$ without prior knowledge of the number of families, we propose a selection method which combines Ward's algorithm, a general agglomerative hierarchical clustering procedure proposed by \citet*{Ward1963Hierarchical}, with the Sargan test of overidentifying restrictions. Our selection algorithm has two parts. 

The first part is Ward's algorithm, as described in Algorithm \ref{algo:ward} below. 
The algorithm aims to minimize the total within-cluster sum of squared error. 
This is achieved by minimizing the increase in within-cluster sum of squared error at each step. The method generates a path of cluster assignments with $K$ clusters at each step such that $K \in \{1,...,\binom{J}{P}\}$. After obtaining the clusters for each $K$, we use a downward testing procedure based on the Sargan-test to select the set of valid IVs (Algorithm \ref{algo:Sargan}). 
Ward's Algorithm works as follows 
\begin{algorithm}\label{algo:ward}
Ward's algorithm
\begin{enumerate}
	\item \textbf{Input:} Each just-identified point estimate is calculated. The Euclidean distance between all of these estimates is calculated and written as a dissimilarity matrix.
	\item \textbf{Initialization:} Each just-identified estimate has its own cluster. The total number of clusters in the beginning hence is $\binom{J}{P}$.
	\item \textbf{Joining:} The two clusters which are closest as measured by their weighted squared Euclidean distance $\frac{|\mathcal{S}_k||\mathcal{S}_l|}{|\mathcal{S}_k| + |\mathcal{S}_l|}||\bar{\mathcal{S}}_k - \bar{\mathcal{S}}_l||^2$ are joined to a new cluster. $|\mathcal{S}_k|$ is the number of estimates in cluster $k$. $\bar{\mathcal{S}}_k$ denotes the mean of cluster $k$, which is the arithmetic mean of all the just-identified estimates in $\mathcal{S}_k$.
	\item \textbf{Iteration:} Step 3 is repeated until all just-identified estimates are in one cluster.
\end{enumerate}
\end{algorithm}
\noindent 
This yields a path of $S = \binom{J}{P}-1$ steps, on which there are clusters of size $K \in \{1, ..., \binom{J}{P}\}$. 
\citet{Ward1963Hierarchical} also allows for alternative objective functions, which are associated with different dissimilarity metrics. Our motivation for using the Euclidean distance is that the objective function is the intra-cluster variance or the sum of residual sum of squares. We discuss alternative choices of these so-called linkage methods and dissimilarity metrics in Appendix \ref{sec:ProximityMeasures}. 
The downward testing procedure considers the selection made in Algorithm 1 for each number of clusters $K \in \{1,...,\binom{J}{P}-1\}$, and chooses the smallest $K$ such that the selected group passes the Sargan test:
\begin{algorithm}\label{algo:Sargan}
Downward testing procedure
\begin{enumerate}
	\item Starting from $K = 1$, find the cluster that contains the largest number of just-identified estimators. In the first step, all estimators are in one cluster. 
	\item Do Sargan test on the IVs associated with the largest cluster, using the rest as controls. If there are multiple such clusters, select the one with the smallest Sargan statistic. 
	\item Repeat the procedure for each $K = 2,..., \binom{J}{P}-1$.
	\item Stop when for the first time the model selected by the largest cluster at some $K$ does not get rejected by the Sargan test. 
	\item Select the instruments associated with the cluster from Step 4 as valid instruments.
\end{enumerate}
\end{algorithm}
\noindent 
The Sargan statistic in Step 4 is given by
\begin{equation*}
Sar\left(K\right)=\frac{\widehat{\mathbf{u}}(\widehat{\bm{\theta}}_K)^{\prime}\mathbf{Z}\left(\mathbf{Z}^{\prime}\mathbf{Z}\right)^{-1}\mathbf{Z}^{\prime}\widehat{\mathbf{u}}(\widehat{\bm{\theta}}_K)}{\widehat{\mathbf{u}}(\widehat{\bm{\theta}}_K)^{\prime}\widehat{\mathbf{u}}(\widehat{\bm{\theta}}_K)/n}
\end{equation*}
where $\hat{\bm{\theta}}_K$ is the 2SLS estimator using the IVs associated with the largest cluster for each $K$ as valid and controlling for the rest, and $\widehat{\mathbf{u}}(\widehat{\bm{\theta}}_K)$ is the 2SLS residual. We show later that to guarantee consistent selection, the critical value for the Sargan
test, denoted by $\xi_{n, J - |\hat{\mathcal{I}}| - P}$ should satisfy $\xi_{n, J - |\hat{\mathcal{I}}| - P} \rightarrow \infty$ and $\xi_{n, J - |\hat{\mathcal{I}}| - P} = o(n)$. In practice, we choose the significance level $\frac{0.1}{log(n)}$ following \citet{Windmeijer2021Confidence}.\footnote{This significance level has been suggested by \citet{Belloni2012Sparse}.} The downward testing procedure can also be described as
\begin{equation}\label{eq:SelectedDownward}
	\hat{\mathcal{V}}^{dts} = \{\hat{\mathcal{V}}^{Sar}(K), \, K= min(1, ..., \binom{J}{P}-1): Sar(\hat{\mathcal{V}}^{Sar}(K)) < \xi_{n, J - |\hat{\mathcal{I}}| - P}. \}
\end{equation}

\begin{center}
\textbf{[Figure 1 here]}
\end{center}

\begin{figure}[h]
	\caption{Illustration of the Algorithm with Two Regressors \label{fig:Ward2}}
	\includegraphics[scale=0.37, trim=95 220 95 195, clip]{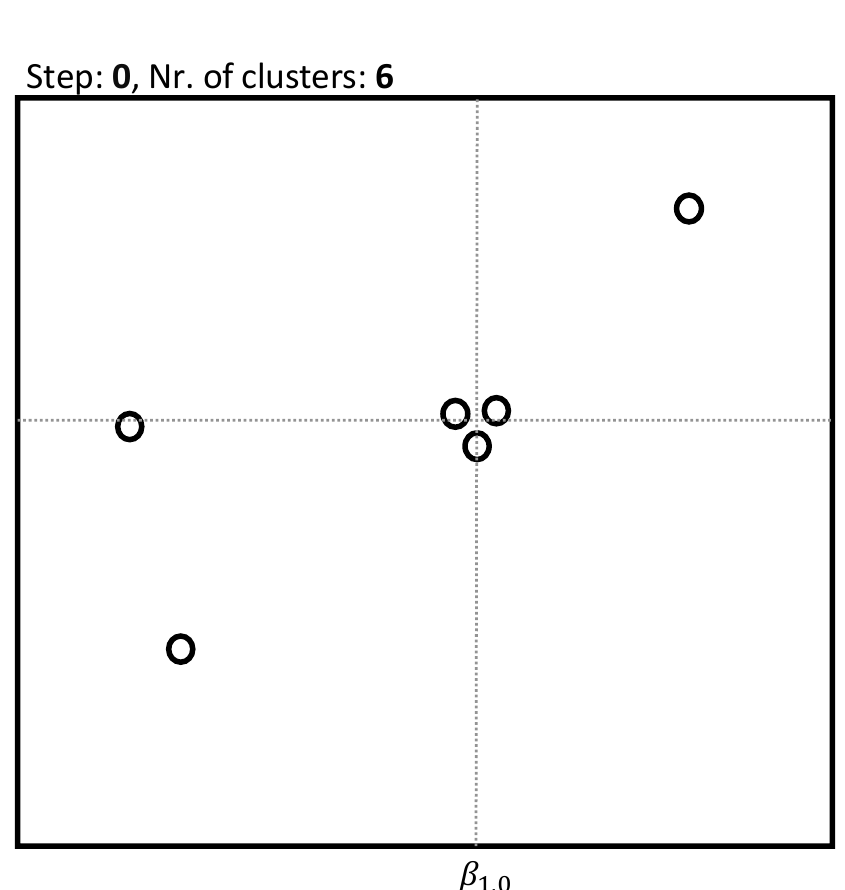}
	\includegraphics[scale=0.37, trim=95 220 95 195, clip]{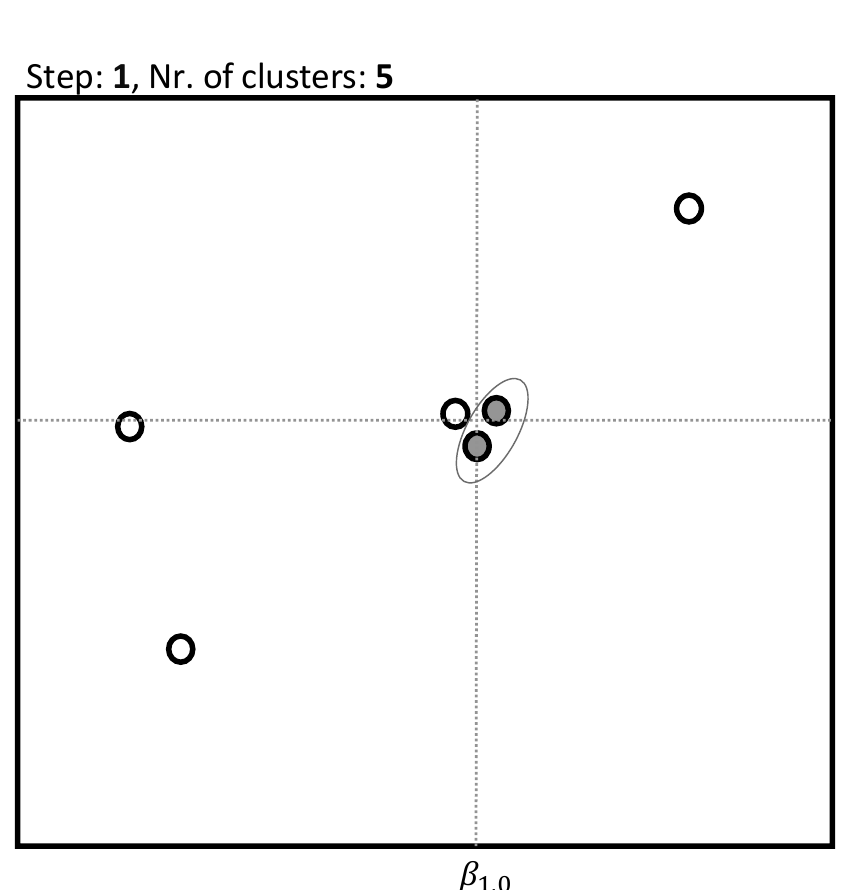}
	\includegraphics[scale=0.37, trim=95 220 95 195, clip]{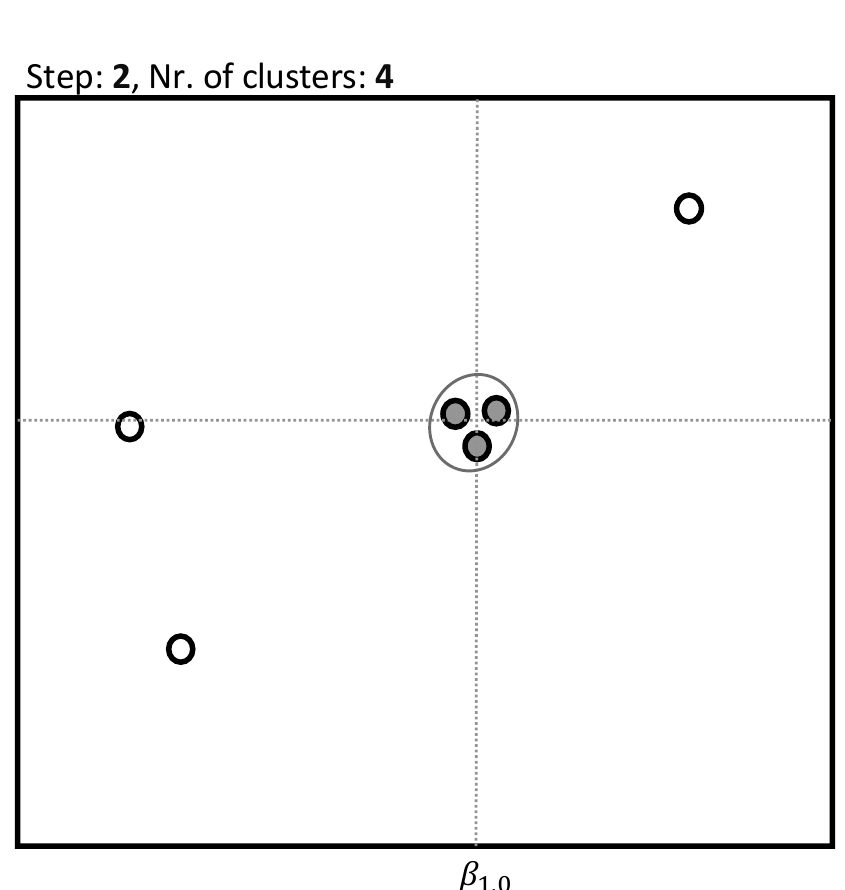}
	
	\includegraphics[scale=0.37, trim=95 220 95 195, clip]{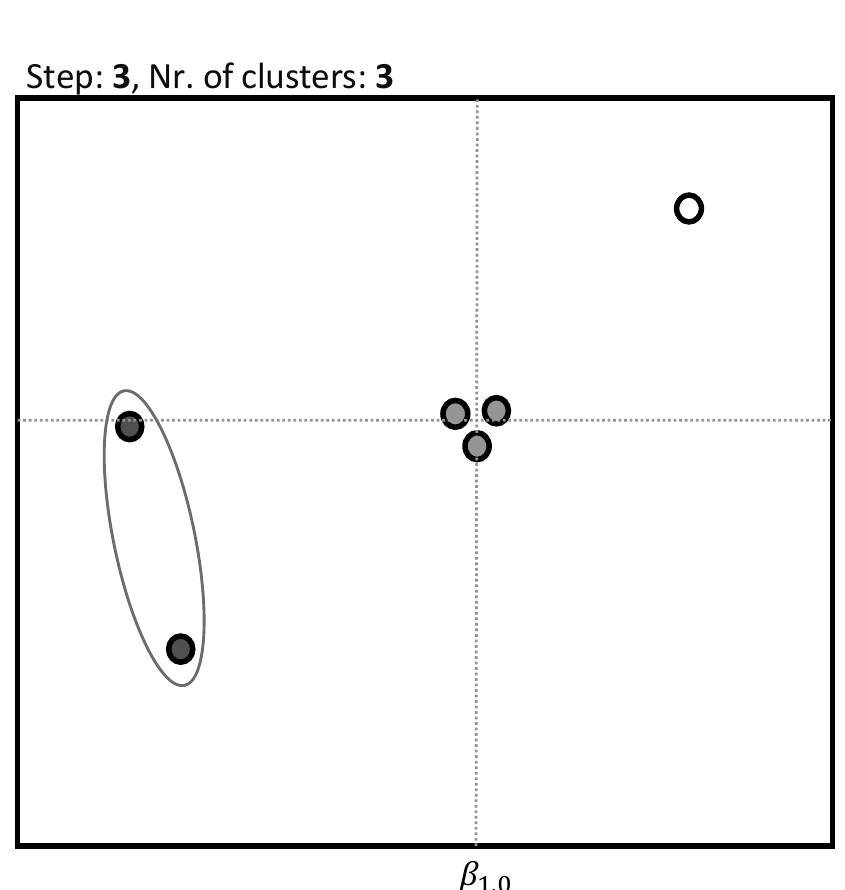}
	\includegraphics[scale=0.37, trim=95 220 95 195, clip]{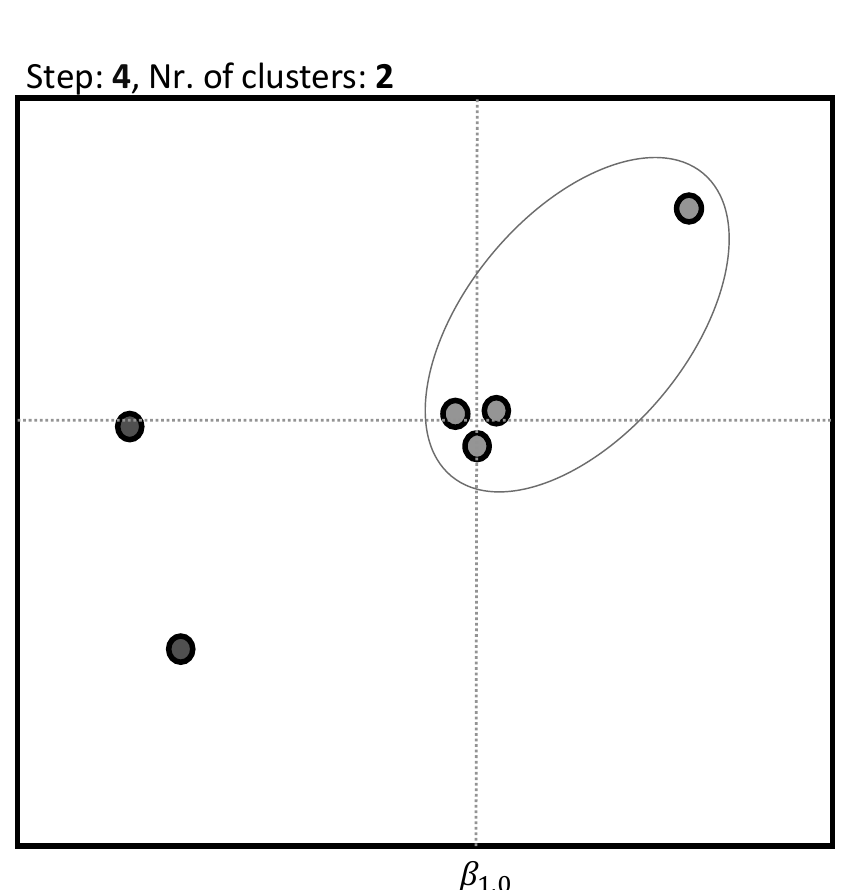}
	\includegraphics[scale=0.37, trim=95 220 95 195, clip]{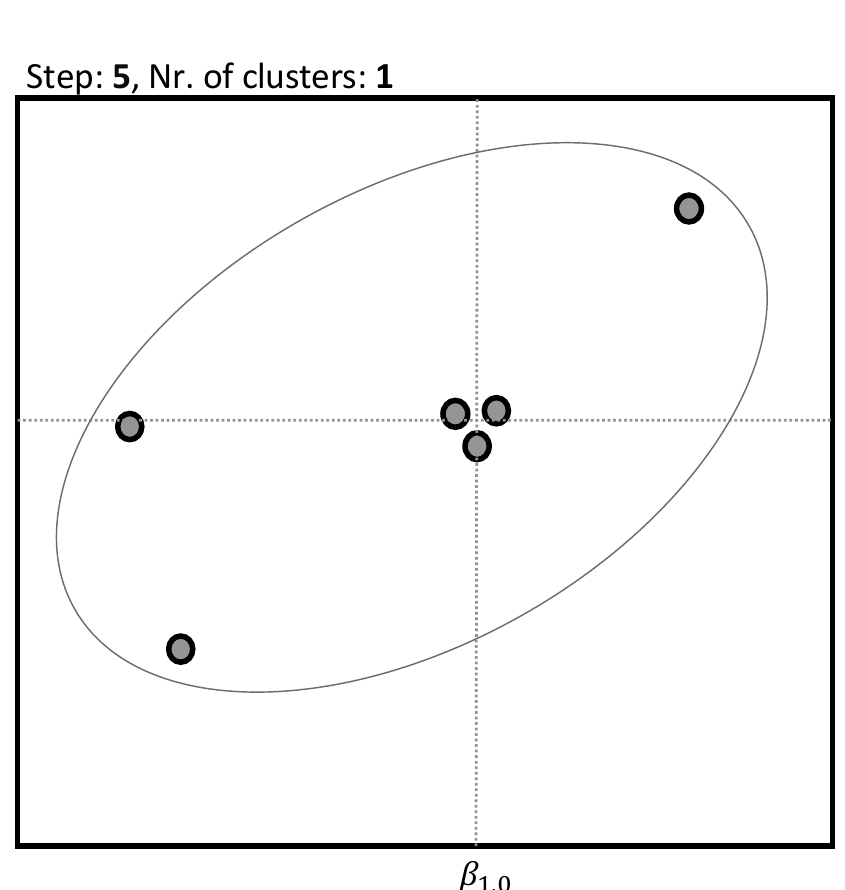}
\end{figure}

\noindent In figure \ref{fig:Ward2}, the procedure is illustrated for a situation with four IVs and two endogenous regressors. Instrument No. 1 is invalid, because it is directly correlated with the outcome, while the remaining three IVs (2, 3, 4) are related with the outcome only through the endogenous regressors and are hence valid. 
In the first graph on the top left, we have plotted each just-identified estimate. The horizontal and vertical axes represent coefficient estimates of the effects of the first ($\beta_1$) and second regressor ($\beta_2$), respectively. Each point has been estimated with two IVs, in this case with IV pairs 1-2, 1-3, 1-4, 2-3, 2-4 and 3-4, because there are four candidate IVs.

In the initial Step (0), each just-identified estimate has its own cluster. In step 1, we join the estimates which are closest in terms of their Euclidean distance, e.g. those estimated with pairs 2-3 and 2-4. These two estimates now form one cluster and we only have five clusters. We re-estimate the distances to this new cluster and continue with the procedure, until there is only one cluster left in the bottom right graph. We continue with Algorithm \ref{algo:Sargan} and evaluate the Sargan test at each step, using the IVs contained in the largest cluster. When the p-value is larger than a certain threshold, say $0.1/log(n)$, we stop the procedure and select the corresponding cluster as the set of valid IVs. Ideally this will be the case at step 2 or 3 of the algorithm, because here the largest cluster (in light gray) is formed only by valid IVs (2,3 and 4). If this is the case, only the valid IVs are selected as valid.

To make the procedure robust to heteroskedasticity, clustering and serial correlation, the Sargan test can be replaced with a robust score test, such as the Hansen J-test \citep*{Hansen1982Large}, analogously to \citet{Windmeijer2021Confidence}.

\subsection{Oracle Selection and Estimation Property}\label{sec:oracle}
Next, we state the theoretical properties of the selection results obtained by Algorithm \ref{algo:ward} and Algorithm \ref{algo:Sargan} and the post-selection estimators. 
We establish that our method achieves oracle properties in the sense that it selects the valid IVs consistently, and that the post-selection estimator has the same limiting distribution as if we knew the true set of valid IVs. 

\begin{theorem}{Consistent selection\label{th:ConsistentSelection}}\\
Let $\xi_n$ be the critical value for the Sargan test in Algorithm 2. Let $\hat{\mathcal{V}}^{dts}$ be the set of IVs selected from Algorithms \ref{algo:ward} and \ref{algo:Sargan}. Under Assumptions 1 - 6, for $\xi_n \rightarrow \infty$ and $\xi_n = o(n)$, 
$$\underset{n \rightarrow \infty}{lim} P (\hat{\mathcal{V}}^{dts} = \mathcal{V}) = 1.$$
\end{theorem}

\noindent 
The post-selection 2SLS estimator using the selected valid instruments and controlling for the selected invalid IVs has the same asymptotic distribution as the oracle estimator:

\begin{theorem}{Asymptotic oracle distribution\label{th:AsymptoticOracleDistribution}}\\
Let $\mathbf{Z}_{\hat{\mathcal{I}}} = \mathbf{Z} \setminus \mathbf{Z}_{\hat{\mathcal{V}}^{dts} }$ with $\mathbf{Z}_{\hat{\mathcal{I}}}$ ,  $\mathbf{Z}_{\hat{\mathcal{V}}^{dts} } $ being the selected invalid and valid instruments respectively. Let $\hat{\bm{\beta}}_{\hat{\mathcal{V}}^{dts}} $ be the 2SLS estimator given by
$$\hat{\bm{\beta}}_{\hat{\mathcal{V}}^{dts}} = (\hat{\mathbf{D}}^\prime \mathbf{M_{\mathbf{Z}_{\hat{\mathcal{I}}}}}\hat{\mathbf{D}})^{-1}\hat{\mathbf{D}}^\prime \mathbf{M_{\mathbf{Z}_{\hat{\mathcal{I}}}}} \mathbf{y}$$
Under Assumptions 1-6, the limiting distribution of $\hat{\bm{\beta}}_{\hat{\mathcal{V}}^{dts}} $ is
\begin{equation*}
    \sqrt{n} (\hat{\bm{\beta}}_{\hat{\mathcal{V}}^{dts}}  - \bm{\beta}) \overset{d}{\rightarrow} N(0, \bm{\Sigma}^2_{or})
\end{equation*}
where $\bm{\Sigma}_{or}^{2}$ is the asymptotic variance for the oracle 2SLS estimator given by
\begin{equation*}\label{sig2bor}
\bm{\Sigma}_{or}^{2} = \sigma_u^2(\mathbf{Q}_{ZD}^{\prime}\mathbf{Q}_{ZZ}^{-1}\mathbf{Q}_{ZD}-\mathbf{Q}_{Z_\mathcal{I}D}^\prime\mathbf{Q}_{Z_\mathcal{I}Z_\mathcal{I}}^{-1}\mathbf{Q}_{Z_\mathcal{I}D})^{-1}
\end{equation*}
with $\mathcal{I}$ being the true set of invalid instruments. 
\end{theorem}

\noindent
The proof of Theorem \ref{th:AsymptoticOracleDistribution} follows from the proof of Theorem 2 in \citet{Guo2018Confidence}. 

\section{Monte Carlo Simulations}\label{sec:MonteCarlo}
\subsection{Strong Instruments}\label{sec:strongsimulation}
We conduct three sets of Monte Carlo simulations to illustrate the performance of our method under different settings. First we evaluate the method in the setup where all the candidate instruments are strong instruments. We then relax Assumption \ref{ass:FirstStageSingle} such that there are some weak instruments among the candidates. After that we inspect the methods for the case with local-to-zero violations and check the performance of the method when ruling out globally and locally invalid IVs. Here we present the case of multiple regressors. For the single regressor designs, see Appendix \ref{sec:SimulationsAppendix}, where we also compare with the performance of HT and CIM.

We follow the setting in \citet{Windmeijer2021Confidence}: There are 21 candidate instruments, 12 of which are invalid, while 9 are valid with $\bm{\alpha} =  \left( \bm{\iota}_6', \, \, \, 0.5 \bm{\iota}_6', \,\,\, \bm{0}_9' \right)'$ where $\bm{0}_r$ is an $r \times 1$ vector of zeros and $\bm{\iota}_r$ is an $r \times 1$ vector of ones. The first-stage parameters are drawn from uniform distributions as $\bm{\gamma}_1 = unif(1,2)$ and $\bm{\gamma}_2 = unif(3,4)$ when there are two endogenous regressors ($P = 2$), and additionally with $\bm{\gamma}_3 = unif(5,6)$ when there are three regressors ($P = 3$). We set the true treatment effects as $\bm{\beta} = \bm{0}$ and the candidate IVs are generated as $\mathbf{z}_i \sim N(0, \bm{\Sigma_z})$ with $\bm{\Sigma_{z,jk}} = 0.5^{|j-k|}$. Errors are generated from
$$\left(\begin{array}{r}
	u_i \\
	\varepsilon_i
\end{array}\right) \sim N\left(\bm{0}, 
\left(
\begin{array}{cc}
	1 & 0.25\\
	0.25 & 1\\
\end{array}\right) \right) \text{.}
$$
We report the median absolute error (MAE) and the standard deviation (SD) of the IV estimators, and the coverage rate of the $95\%$ confidence intervals (\textit{Coverage}). For IV selection results we report three statistics: the number of selected invalid instruments (\textit{\# invalid}), the frequency of selecting all invalid instruments as invalid (\textit{p allinv}) and the frequency of selecting the oracle model (\textit{p oracle}). We report the results for the oracle model which uses the true set of valid instruments for IV estimation (\textit{Oracle}), the naive model which treats all candidate instruments as valid (\textit{Naive}), and the AHC method (\textit{AHC}). We run the simulations with number of observations $n\in \{500, 1000, 5000\}$. All statistics are calculated based on $1000$ Monte Carlo replications.

\begin{table}[!htbp] \centering 
  \caption{Simulation Results with More Than One Regressor} 
  \label{tab:MultipleRegressors} 
\begin{tabular}{@{\extracolsep{5pt}} C{0.11\textwidth} C{0.11\textwidth}C{0.11\textwidth}C{0.11\textwidth}C{0.11\textwidth}C{0.11\textwidth}C{0.11\textwidth}} 
\toprule
 & MAE & SD & \# invalid & p allinv & Coverage & p oracle \\ 
\midrule
\textbf{P=2}    &     \multicolumn{6}{c}{\textbf{n=500}}  \\ 
Oracle & 0.049 & 0.085 & 12     & 1    & 0.965   & 1 \\ 
Naive  & 0.597 & 0.377 & 0      & 0    & 0.032   & 0 \\ 
AC     & 0.080 & 0.583 & 12.215 & 0.930& 0.879   & 0.750 \\ 
\midrule
 &   \multicolumn{6}{c}{\textbf{n=1000}}  \\ 
Oracle & 0.044 & 0.068 & 12     & 1    & 0.952   & 1 \\ 
Naive  & 0.658 & 0.272 & 0      & 0    & 0       & 0 \\ 
AC     & 0.055 & 0.343 & 12.202 &0.982& 0.919   & 0.827 \\ 
\midrule
 &     \multicolumn{6}{c}{\textbf{n=5000}} \\ 
Oracle & 0.021 & 0.033 & 12     & 1    & 0.949   & 1 \\ 
Naive  & 0.755 & 0.138 & 0      & 0    & 0       & 0 \\ 
AC     & 0.024 & 0.037 & 12.109 & 1& 0.938   & 0.909 \\ 
\midrule
\textbf{P=3} &\multicolumn{6}{c}{\textbf{n=500}}  \\ 
Oracle & 0.063 & 0.099 & 12     & 1    & 0.952   & 1 \\ 
Naive  & 0.880 & 0.372 & 0      & 0    & 0.002   & 0 \\ 
AC     & 0.121 & 0.804 & 12.190 & 0.794& 0.725   & 0.520 \\ 
\midrule
&    \multicolumn{6}{c}{\textbf{n=1000}} \\ 
Oracle & 0.050 & 0.078 & 12     & 1    & 0.934   & 1 \\ 
Naive  & 0.915 & 0.279 & 0      & 0    & 0       & 0 \\ 
AC     & 0.073 & 0.416 & 12.367 & 0.948& 0.844   & 0.696 \\ 
\midrule
&  \multicolumn{6}{c}{\textbf{n=5000}} \\ 
Oracle & 0.037 & 0.058 & 12     & 1    & 0.919   & 1 \\ 
Naive  & 0.941 & 0.211 & 0      & 0    & 0      & 0 \\ 
AC     & 0.049 & 0.307 & 12.261 & 0.976& 0.853   & 0.797 \\ 
\bottomrule&&&&&& \\[-1.8ex] 
\multicolumn{7}{c}{\begin{minipage}{\textwidth} \footnotesize 
                This table reports median absolute error, standard deviation, 
                number of IVs selected as invalid, frequency with which all invalid 
                IVs have been selected as invalid, coverage rate of the 95 \% confidence interval 
                and frequency with which oracle model has been selected. For the first two, means over the statistic for each regressor are taken.
                The true coefficient is $\bm{\beta}=\bm{0}$. Settings are described in the text. 
                1000 repetitions per setting.
                \end{minipage}} \\ 
\end{tabular} 
\end{table} 

\noindent The results are shown in Table \ref{tab:MultipleRegressors}. The IV selection performance of our method approaches the ground truth as the sample size grows large. This can be seen as \textit{\# invalid} is close to 12, \textit{p allinv} and \textit{p oracle} approach 1 and coverage increases as the sample size grows.
But as the number of endogenous variables increases from 2 to 3, it needs a larger sample size to achieve oracle selection. When $P=3$ and $n=5000$, coverage is still not at its nominal level and \textit{p oracle} is only at about 0.8. 
Generally, the estimation bias of our method approaches the oracle bias and is only a fraction of that of the naive estimator. The existing IV selection methods, HT and CIM, do not allow for multiple regressors and we compare the performance of AHC with these two methods for the single regressor case in Appendix \ref{sec:SimulationsOneRegressor}. We find that our method works as well as the CIM in terms of selection and bias and outperforms HT when the sample size is relatively small.


\subsection{Some Weak Instruments Among the Candidate Instruments\label{sec:weaksimulations}}
Now we examine the performance of AHC when Assumption \ref{ass:FirstStageSingle} is violated, i.e. there are weak instruments among the candidates. 
We focus on the case with two endogenous variables in large samples (fixed sample size $n = 5000$). 
For individually weak instruments, we consider the local to zero setup and set their first stage parameters as $\gamma_j = C/\sqrt{n}$ with $C = 0.1$.
For ease of illustration, simulations are conducted in four simplified designs with $9$ candidate instruments (see Table \ref{tab:weakdoublesetup}).
In Design 1, each IV is valid but only strong for one of the regressors. We are interested to see if the AHC method can include all the IVs as valid. In Design 2, still all the candidate IVs are strong for only one regressor, but some of them are invalid. In the last design, we make some of the IVs weak for both variables and we set some of them to be invalid. 
All other parameters are the same as in Section \ref{sec:strongsimulation}. 

Results are presented in Table \ref{tab:weakdouble}, where we report three additional IV selection statistics: the frequency of selecting all valid and strong instruments as valid (\textit{strongval}), the frequency of selecting all weak invalid instruments as invalid (\textit{weakin}), and the frequency of selecting all weak valid instruments as invalid (\textit{weakva}). In these designs, the oracle model includes only the strong and valid instruments as valid. 

Our primary focus is the selection of all the invalid instruments, especially the weak invalid instruments, as the weak IVs can amplify the bias caused by invalidity (see discussion in Appendix \ref{sec:SimulationsP1Weak}). In all designs, AHC has high frequencies of including all invalid instruments, either weak or strong, as invalid (\textit{p allinv} and \textit{weakin} close to $1$). Hence the MAEs are very similar to those of the oracle. In terms of valid IVs, all the strong valid instruments are included as valid with high frequencies (\textit{strongval} close to 1). However, for weak valid instruments, some of them are still selected as valid, which can be seen from the relatively low \textit{weakva} in Design 3. This may be the main drawback of AHC in the presence of weak IVs. In Design 3, we should start to see the limitations of AHC in the weak IV case, because there are only four just-identifying combinations of strong and valid IVs (1-8, 1-9, 2-8, 2-9), and the remaining 32 combinations either violate assumption \ref{ass:FirstStageSingle} or validity. Still, the MAE and the probability with which all invalid (0.929), all strong, valid (0.904) and especially all weak invalid (0.997) are selected correctly is high. However, the few selection mistakes that occur create a very large variance, that can be seen from the high SD. This is in line with the intuition that bias is amplified in presence of weak and invalid IVs. In the simulations very few outlier estimates can strongly increase the SD. 
In Appendix \ref{sec:SimulationsP1Weak} we compare the AHC method with HT and CIM in presence of weak IVs with $P=1$. We find that AHC outperforms CIM in all settings and it outperforms HT in the case where the valid and strong group is not strictly the largest.

\begin{table}[!t]
    \centering
        \caption{Weak IV Simulation Designs with Two Endogenous Regressors}
    \label{tab:weakdoublesetup}
    \begin{tabular}{cccc|cccc|cccc}
    \toprule
\multicolumn{4}{c}{Design 1} & \multicolumn{4}{c}{Design 2} & \multicolumn{4}{c}{Design 3}\\
IV & $\gamma_1$                              &  $\gamma_2$               & $\alpha$&
IV & $\gamma_1$                              &  $\gamma_2$               & $\alpha$&
IV & $\gamma_1$                              &  $\gamma_2$               & $\alpha$ \\
         \midrule
    $\mathbf{z}_1$& $1$                             &  $C/\sqrt{n}$            &   0&
    $\mathbf{z}_1$& $1$                             &  $C/\sqrt{n}$            &   1&
    $\mathbf{z}_1$& $1$                             &  $C/\sqrt{n}$            &   0\\
    $\mathbf{z}_2$& $2$                             &  $C/\sqrt{n}$           &   0&
    $\mathbf{z}_2$& $2$                             &  $C/\sqrt{n}$           &   1&
    $\mathbf{z}_2$& $2$                             &  $C/\sqrt{n}$            &   0\\
    $\mathbf{z}_3$& $3$                             &  $C/\sqrt{n}$           &   0&
    $\mathbf{z}_3$& $3$                             &  $C/\sqrt{n}$           &   1&
    $\mathbf{z}_3$& $3$                             &  $C/\sqrt{n}$            &   1\\
    $\mathbf{z}_4$& $4$                             &  $C/\sqrt{n}$           &   0 &
    $\mathbf{z}_4$& $4$                             &  $C/\sqrt{n}$           &   0&
    $\mathbf{z}_4$& $C/\sqrt{n}$                   &  $C/\sqrt{n}$            &   1\\
    $\mathbf{z}_5$& $C/\sqrt{n}$                           &  $unif(1,2)$              &   0&
    $\mathbf{z}_5$& $C/\sqrt{n}$                           &  $unif(1,2)$              &   0&
    $\mathbf{z}_5$& $C/\sqrt{n}$                           &  $C/\sqrt{n}$            &   1\\
    $\mathbf{z}_6$& $C/\sqrt{n}$                           &  $unif(1,2)$              &   0&
    $\mathbf{z}_6$& $C/\sqrt{n}$                           &  $unif(1,2)$              &   0&
    $\mathbf{z}_6$& $C/\sqrt{n}$                           &  $C/\sqrt{n}$            &   0\\
    $\mathbf{z}_7$& $C/\sqrt{n}$                           &  $unif(1,2)$              &   0&
    $\mathbf{z}_7$& $C/\sqrt{n}$                           &  $unif(1,2)$              &   0&
    $\mathbf{z}_7$& $C/\sqrt{n}$                           &  $unif(3,4)$              &   1\\
    $\mathbf{z}_8$& $C/\sqrt{n}$                           &  $unif(1,2)$              &   0&
    $\mathbf{z}_8$& $C/\sqrt{n}$                           &  $unif(1,2)$              &   1&
    $\mathbf{z}_8$& $C/\sqrt{n}$                           &  $unif(3,4)$              &   0\\
    $\mathbf{z}_9$& $C/\sqrt{n}$                           &  $unif(1,2)$              &   0&
    $\mathbf{z}_9$& $C/\sqrt{n}$                           &  $unif(1,2)$              &   1&
     $\mathbf{z}_9$& $C/\sqrt{n}$                           &  $unif(3,4)$             &   0\\
    \bottomrule
    \end{tabular}
\end{table}
\begin{table}[!ht]
	\small
	\caption{Some Weak Instruments with Two Endogenous Regressors} 
	\label{tab:weakdouble} 
	\begin{tabular}{@{\extracolsep{5pt}} C{0.07\textwidth}C{0.06\textwidth}C{0.07\textwidth}C{0.1\textwidth}C{0.08\textwidth}C{0.1\textwidth}C{0.08\textwidth}C{0.08\textwidth}C{0.09\textwidth}} 
		&&&&&&&&\\[-1.8ex]
		\toprule 
		& MAE & SD &\# invalid & p allinv & strongval & weakin & weakva & coverage \\ 
		\midrule 
		\multicolumn{9}{c}{\textbf{Design 1}} \\ 
		oracle   & 0.004 & 0.006 & 0     & 1 & 1     &-  &-& 0.894  \\
		naive & 0.004 & 0.006 & 0 & 1 & 1 & - & - & 0.894\\
		AHC      & 0.004 & 0.006 & 0.020 & 1 & 0.988 &-  &-& 0.887  \\ 
		\midrule
		\multicolumn{9}{c}{\textbf{Design 2}} \\ 
		oracle   & 0.008 & 0.011 & 5     & 1 & 1     &-  &-& 0.895  \\ 
		naive & 0.571 & 0.013 & 0 & 0 &1 & - & - & 0\\
		AHC      & 0.009 & 0.301 & 5.004 & 0.811 & 0.810 &-  &-& 0.723  \\ 
		\midrule
		\multicolumn{9}{c}{\textbf{Design 3}} \\ 
		oracle   & 0.010 & 0.015 & 5     & 1 & 1     &1  &1& 0.908  \\ 
		naive & 0.730 & 0.021 & 0 & 1 & 1 & 0 & 0 & 0\\
		AHC      & 0.010 & 11.171 & 4.215 & 0.929 & 0.904 &0.997  &0.122& 0.863 \\ 	
	\bottomrule
	&&&&&&&&\\[-1.8ex]
		\multicolumn{9}{c}{\begin{minipage}{\textwidth} \footnotesize 
				This table reports median absolute error, standard deviation,
				number of IVs selected as invalid, frequency of all invalid 
				IVs selected as invalid, frequency of all valid and strong instruments selected as valid, frequency of all weak invalid instruments selected as invalid, frequency of all weak valid instruments as invalid, and coverage rate of the 95\% confidence interval. 1000 repetitions per setting.
		\end{minipage}} \\ 
	\end{tabular} 
\end{table}


\subsection{Simulations with Local Violations}\label{sec:simlocalviolations}
The results of this paper and also of \citet{Guo2018Confidence} and \citet{Windmeijer2021Confidence} are pointwise: if we take the values of the violations $\bm{\alpha}$ as fixed, the convergence results hold. However, \citet{Leeb2005Model} point out that for any $n$ there can be DGPs such that the asymptotical results regarding selection and inference do not hold. For example, when the violations of the exclusion restriction are local-to-zero, selection algorithms might select invalid instruments with positive probability leading to an asymptotic bias. 
To illustrate this problem we follow the literature on near exogeneity as in \citet{Newey1985Generalized}, \citet[CHR]{Conley2012Plausibly}, \citet*{Caner2014Near} and \citet*{Andrews2017Measuring} we consider a setting in which the violations consist of a \textit{global} part $\alpha_j$ and a \textit{local} part $\tau_j$, collected in $\bm{\tau}$, that disappears asymptotically, with $\tau_j \rightarrow 0$ as $n \rightarrow \infty$. The model now is 
\begin{equation}\label{eq:StructuralLocalViolations}
	\mathbf{y} = \mathbf{D}\bm{\beta} + \mathbf{Z}_{\mathcal{I}}\bm{\alpha}_{\mathcal{I}} + \mathbf{Z}\bm{\tau}  + \mathbf{u} \textbf{,}
\end{equation}
\noindent where both $\alpha_j$ and $\tau_j$ can differ by IV. As in \citet{Caner2014Near}, we model $$\tau_j = \frac{c}{n^{\kappa}}\textbf{,}$$ with $\kappa > 0$. Here, $\kappa = 1/2$ is a mild violation of an exclusion restriction, $1/2 < \kappa < \infty$ is the range of minor and $0< \kappa < 1/2$ is the range of strong violations. 
Because our testing procedure builds on the Sargan test, we investigate its performance when IVs from the same group and mixtures of different groups are tested, with varying degree of violations. This helps us formulate expectations on how the testing procedure will perform under various DGPs. 
The results on the Sargan test, which are summarized in Table \ref{tab:Summary-Near} in the Appendix suggest that with no and minor local violations the downward testing procedure will accept for groups and reject for mixtures, which is desirable. With strong violations, we expect that the procedure detects strong local violations for certain configurations of $\xi$ (critical value of the Sargan test) and $\kappa$. 
We expect the most problematic case to be the one with mild violations, because the test accepts when globally valid IVs have mild local violations, leading to bias. However, this is a general shortcoming of 2SLS. The selection method can help discern globally invalid and strong local violations from globally valid and minor local violations. This has the potential to reduce bias. Therefore, as further discussed in the Appendix, in this type of setting our method can be seen as a pre-screening procedure to weed out global violations and local but strong violations to reduce bias. 

To illustrate the above discussion, we conduct simulations to investigate the performance of the methods when there are local violations. 
Our primary focus is to exclude global and local strong violations. Consider a setting with $P=2$, $n=5000$, $\bm{\alpha} = \left( \bm{0}_6', \,\,\bm{\iota}_6 0.5', \,\, \bm{0}_9'\right)'$ and $\bm{\tau} = \left(\bm{\iota}_{12}'c_\alpha/n^{k}, \,\, \bm{0}_{9}'\right)'$ where $c_\alpha = 2$. All other parameters are chosen in the same way as in the preceding section, in Table \ref{tab:MultipleRegressors}. We then vary $\kappa$ in three designs:

\begin{itemize}
	\item Design 1, minor violations, $\kappa = 3/4$
	\item Design 2, mild violations, $\kappa=1/2$
	\item Design 3, strong violations, $\kappa=1/4$.
\end{itemize}
We report two additional statistics: the number of invalid instruments with global violations selected as invalid (\textit{global violation}) and the frequency of selecting all the invalid instruments with global violations as invalid (\textit{p allins}). We report the performance of the oracle estimator which treats the globally but not the locally invalid IVs as invalid (\textit{oracle global}) and the estimator that treats both globally and locally invalid as invalid (\textit{oracle}). The simulation results are presented in Table \ref{tab:localviolationmulti}. 
\afterpage{
\begin{table}[!htbp] \centering 
	\caption{Local Violations with Two Regressors} 
	\label{tab:localviolationmulti} 
        \begin{tabular}{lC{0.08\textwidth}C{0.08\textwidth}C{0.12\textwidth}C{0.09\textwidth}C{0.12\textwidth}C{0.09\textwidth}C{0.1\textwidth}} 
		\toprule
		& MAE &SD & \# invalid &    p allinv & global viol &p allins   & coverage \\ 
		\midrule
            \multicolumn{8}{c}{\noindent\small\textbf{Design 1}: $\kappa = 3/4$} \\
		oracle          & 0.021  &0.032  & 12     & 1     & 6     & 1       &0.931\\ 
		oracle global   & 0.013  &0.019  & 6      & 0.5   & 6     & 1       &0.927\\ 
		naive           & 0.143  &0.060  & 0      & 0     & 0     & 0       &0.000\\ 
		AHC             & 0.014  &0.021  &6.16   &  0     &6      & 1       &0.919  \\ 
		\midrule
            \multicolumn{8}{c}{\noindent\small\textbf{Design 2}: $\kappa = 1/2$} \\
		oracle          & 0.021  &0.032  & 12     & 1     & 6     & 1     &0.931 \\ 
		oracle global   & 0.020  &0.020  & 6      & 0.5   & 6     & 1     &0.793\\ 
		naive           & 0.138  &0.062  & 0      & 0     & 0     & 0     &0.000\\ 
	        AHC         &0.033   &0.047  &8.623   &0.001  &6      &1      &0.569\\ 
		\midrule
            \multicolumn{8}{c}{\noindent\small\textbf{Design 3}: $\kappa = 1/4$} \\
		oracle          & 0.021   &0.032   & 12     & 1       & 6     & 1       &0.931 \\ 
		oracle global   & 0.139   &0.035   & 6      & 0.5     & 6     & 1       &0.000\\ 
		naive           & 0.139   &0.082   & 0      & 0       & 0     & 0       &0.000\\ 
		AHC             & 0.023   &0.059   &12.457  &0.985    & 6     & 1       &0.893\\ 
		\bottomrule \\[-1.8ex] 
		\multicolumn{8}{c}{\begin{minipage}{\textwidth} \footnotesize 
				This table reports median absolute error, standard deviation, 
				number of IVs selected as invalid, frequency of all invalid 
				IVs selected as invalid, number of invalid IVs with global violations selected as invalid, frequency of selecting all the invalid IVs with global violations as invalid, and coverage rate at 5\% significance level. 1000 repetitions per setting.
		\end{minipage}} 
	\end{tabular} 
\end{table}
}

In Design 1, the \textit{oracle} and \textit{oracle global} both have low bias and variance, while the naive 2SLS is clearly biased. AHC works well in selecting the six global violations and the MAE as well as SD approach the performance of \textit{oracle global}. Keeping IVs with minor violations does not hurt the asymptotic performance of estimators. Coverage is also close to the nominal level. In Design 2, oracle global has a lower coverage probability than the oracle that takes all invalid IVs into account. AHC inherits this problem and displays a larger bias and SD than the oracle estimator and a coverage probability of only 0.55. This confirms that in the case with mild violations in fact our method can underperform, but in this setting it still outperforms the naive 2SLS which has coverage 0 and higher bias. With strong violations in Design 3, the oracle estimator that does not take local violations into account performs almost as poorly as the naive estimator and hence it is important to also select the IVs with local violations. However, AHC selects all invalid IVs as invalid 98.6\% of the time and coverage is at 0.894, not far from the nominal level. Overall, these results are in line with what was suggested by our results on the Sargan test. 

A recent approach to address local violations is \citeS{Guo2023Causal} searching and sampling method into which AHC can be incorporated. For this, an algorithm searches over values of $\beta$ and categorizes IVs into valid and invalid. When a plurality of IVs is chosen as valid the value of $\beta$ is assigned to the CI. The entire CI is the union of all $\beta$ values for which a majority is chosen as valid. This approach delivers a uniformly valid confidence interval when more IVs are valid than locally invalid. We use this approach for the single regressor case in Table \ref{tab:SearchAndSampling} in the Appendix, and find that in fact the coverage is larger than the nominal level and CIs are conservative. Another problem with this procedure is that it has been developed for the single regressor case, $P=1$. If one was to extend this procedure to multiple endogenous regressors, it would be necessary to search over values of $\bm{\beta}$ not over the real line but in $\mathbb{R}^P$-space which becomes infeasible quickly. Developing an extension of this method for the multiple regressor case is outside of the scope of this paper and we leave it for future research. 

\section{The Effect of Immigration on Wages}\label{sec:migration}

Many studies\footnote{An overview of the literature can be found in \citet{Dustmann2016impact}.} have tried estimating the contemporaneous effects of immigration on labor market outcomes via some variant of the following equation: \begin{equation}\label{eq:BPcontemp}
	\Delta y_{lt} = \beta^{short} \Delta immi_{l,t} + \psi_{t} + \varepsilon_{lt} \text{,}
\end{equation}

\noindent with year $t$, location $l$, the outcome $\Delta y_{lt}$ being the change in log weekly wages of high-skilled workers. The independent variable is $\Delta immi_{l,t}$, denoting the current change of immigrants in employment. The coefficient of interest is the short-term (contemporaneous) effect $\beta^{short}$.  
Decade fixed-effects are captured by $\psi_{t}$ and $\varepsilon_{lt}$ is the error term. Commuting-zone fixed effects are eliminated through first-differencing as is standard with panel data \citep[][, p. 315]{Wooldridge2010Econometric}. 
The dataset used for estimation is taken from \citet*{Basso2015Association}, who use data from the Census Integrated Public Use Micro Samples and the American Community Survey \citep{Ruggles2015Integrated} with $t \in \{1990, 2000, 2010\}$ and 722 commuting zones.

The key econometric challenge is that migrants choose where to live endogenously, for example, based on economic conditions in a region, creating bias in the estimates. 
A much-used estimation strategy to address this issue is to use historical settlement patterns of migrants from many countries of origin as instruments. When earlier migrants attract migrants at later points in time, the instruments are relevant. This identification strategy dates back to \citet*{Altonji1991effects} and many papers use it \citep[see][]{Jaeger2020Shift}. 
This approach is canonical and highly relevant in applied economics. In a recent paper, \citet{Goldsmith-Pinkham2020Bartik} discuss a class of IVs which are extensively used in labor economics. These so-called shift-share IVs combine the previous settlement shares with aggregate-level shocks, so-called shifts. A sufficient condition for this type of IVs to be valid is that all shares are valid. \citet{Apfel2024Relaxing} discusses the use of IV selection methods for shift-share instrumental variable analysis in this context. 

We use all shares of foreign-born people (migrants, analogously), $s_{jlt_0}$, in working age from a certain origin country $j$ at a base period $t_0$ in region $l$, measured relative to their total number in the country. We use origin-specific shares from 19 origin country groups and base years 1970 and 1980 as separate IVs and obtain $L=38$ IVs. It is usually expected that the reasons that attracted migrants in the past are quasi-random as compared with current migration. Validity is typically defended on these grounds. 

However, these previous settlement patterns might be invalid for various reasons. 
\citet{Jaeger2020Shift} show that IV estimators that rely on this kind of exclusion restriction might be inconsistent, first, because of correlation of the IVs with unobserved demand shocks and, second, because of dynamic adjustment processes: through complementarities between factors of production the initial effect of immigration might propagate in time. For example, native workers might migrate as a response to immigration, leading to a negative shock and hence to a continuation of the initial effect. To capture these dynamic adjustments, they propose to use the following model specification: 

\begin{equation}\label{eq:BP}
	\Delta y_{lt} = \beta^{short} \Delta immi_{l,t} + \beta^{long} \Delta immi_{l,t-10} + \psi_{t} + \varepsilon_{lt} \text{.}
\end{equation}

\noindent where we have included the lagged regressor $\Delta immi_{l,t-10}$ with the coefficient $\beta^{long}$. 
Of course, the lagged regressor will also be subject to the same endogeneity problem as before and hence should also be instrumented. 

The shares of migrants might still be endogenous if $\Delta immi_{l,t-10}$ does not perfectly control for the direct effects of lagged immigration and the instruments themselves still have a direct effect. Moreover, serial correlation in unobserved shocks and correlation between unobservables in the past and shares of immigrants might still be an issue. 

Although some of the shares might be invalid, it is reasonable to believe that the plurality rule holds. The migration literature uses migrant shares as instruments, because it is believed that long lags break the serial correlation in unobservables and all correlation between shares and outcome is mediated through the treatment variable. In our application, half of the shares are from 1970, half from 1980. Typically a 10-year lag is thought to be long enough to break serial correlation in unobservable shocks \citep{Jaeger2020Shift} and shares from 1980 would be used for the construction of the instrument. Still some of the shares might be invalid for various reasons. 
First, if shares from 1980 are correlated with unobservables and these are still serially correlated, shares from 1980 would be invalid. 
Secondly, visa composition might play a role. In the 1970s many migrants were in the US on family visas, while by 1980 many visa were work-related \citep{Jaeger2007}. 
The latter would be more likely to choose their location based on labor market shocks. Migrant groups differ by their visa composition even in the same cohort and those with a high proportion of work-related visa would be more prone to invalidity. We expect these concerns to hold for a few but not all shares. In this sense, plurality is not at all restrictive, especially as compared to the baseline assumption which means to assume that all shares are valid. If the shares from 1970 are all valid and at least one of the shares from 1980 is valid, a majority, not only a plurality assumption would hold. For these reasons, researchers should feel comfortable with the plurality assumption and we can use AHC to select valid and invalid shares and to estimate the effects of immigration.\footnote{These issues have also been similarly discussed in \citet{Apfel2024Relaxing} who uses the same empirical example to illustrate share selection via the adaptive LASSO and the Confidence Interval Method. Notably, \citet{Goldsmith-Pinkham2020Bartik} propose weights that measure how the overall inconsistency of the 2SLS estimator changes in relative terms when specific shares are invalid, specifically in the migration economics context. Robustness checks excluding the top-5 weights are then proposed. This implies that a some valid-some invalid setting is credible in general. However, these weights do not tell us which shares are indeed valid and even shares that are not influential in relative terms can still lead to large absolute bias.} 

\begin{table}[!t]
	\begin{center}	
		\caption{Impact of Immigration\label{tab:BP}} \begin{tabular}{l*{5}{c}}
          &\multicolumn{1}{c}{OLS}&\multicolumn{1}{c}{2SLS}&\multicolumn{1}{c}{OLS}&\multicolumn{1}{c}{2SLS}&\multicolumn{1}{c}{2SLS AHC}\\

\hline $\Delta immi_{lt}$&    0.451&  0.519&    0.586&    0.877&    1.215\\
          & (0.0674)&  (0.173)& (0.0935)&  (0.460)&  (0.505)\\

$\Delta immi_{lt-10}$&   &           &   -0.197&   -0.249&   -0.667\\
          &         &      & (0.0814)&  (0.321)&  (0.403)\\

\hline Nr inv&          &     0&         &        0&        2\\
P-value   &          & .0067&         &    .0126&    .1137\\
CD        &        &64.0901&         &  17.7088&  17.9824\\
\hline \multicolumn{6}{p{.8\textwidth}}{\footnotesize n = 2166 (722 CZ $\times$ 3), L = 38. Robust standard errors in parentheses. Observations weighted by beginning-of-period population. Significance level in testing procedure: 0.013.} \end{tabular}

	\end{center}
\end{table}

The results can be found in Table \ref{tab:BP}. In order to compare our results with \citet{Basso2015Association} we start with the estimate of the short-term effect only. The first column shows results for ordinary least squares: the contemporaneous effect is 0.451, which is significant. For the 2SLS regression in column 2 we use all shares from 1970 and find a significant effect of 0.519. The coefficient estimate in column 3 of the first line of Table 7 in  \citet*{Basso2015Association} is a significant 0.38 and is the one most comparable to our estimate of the short-term effect in the second column. These estimates differ because the authors also use the 1980 data for estimation and we will use shares from this year as IVs in the following, and for comparability with our own results we omit this year from the estimation with a single regressor. Moreover, the authors use a non-linear version of the shift-share IV but motivate the instrument through the use of shares and hence we can use the 2SLS with all shares used as IVs as the baseline estimate.\footnote{\citet{Goldsmith-Pinkham2020Bartik} show that an overidentified GMM regression weighted by the shifts is equivalent to the shift-share IV estimate.} Note that the estimands of SSIV and 2SLS in this case don't have to be identical. The Hansen-Sargan test rejects at the 0.01 significance level and the Cragg-Donald test statistic is at 64. Introducing lagged immigration and estimating via OLS in Column 3, the lagged effect is -0.197 and statistically significant. When using all shares as valid IVs also including those from 1980, both short- and long-term effects are higher in absolute terms but only the contemporaneous effect is marginally statistically significant at the 10 percent level. The Hansen-Sargan test for this model gives a $p$-value of 0.0126, which is lower than the proposed significance level of $0.1/log(n)$.

When using AHC with this significance level in the downward testing procedure, two origin country shares are selected as invalid: the share of Scandinavians and North Europeans in the 1980s and the share of foreign-born from the Baltic States. The coefficient estimates of the short- and long-term effects increase considerably in absolute terms. Now, the coefficient estimate of the short-term effect is statistically significant at the 5 percent level and the estimate of the long-term effect is significant at the 10 percent level. This indicates that the use of AHC indeed makes a big difference. Moreover, the $p$-value of the Sargan test is now 0.1137, exceeding the threshold of $0.1/log(n)=0.013$. The coefficient estimates for short- and long-term effects are considerably higher than OLS and 2SLS estimates. The short-term effect is at 1.215 as compared to the 0.887 estimated by 2SLS and is still almost double the statistically insignificant long-term effect (-0.667). This suggests considerable sensitivity to endogenous shares and is interesting as potentially the positive effects of immigration on wages of the high-skilled are larger than previously thought. 

The two selected shares are similar a priori in that they are from the same region. 
It is plausible that they are indeed invalid, through a combination of two reasons: invalid and weak IVs. As to invalidity, correlation of unobserved shocks might be the culprit. The concentration of Americans of Swedish descent is highest in the Midwest, especially in Minnesota. Cheap land attracted northern European settlers to these agricultural centers. The agricultural sector remained one of the main sectors in this region in subsequent decades, affecting wages at later periods. Similarly, Baltic migrants concentrated in the same large cities which attracted migrants with high wages in the subsequent decades. It is therefore well possible that wages or unobserved productivity shocks that drove initial settlement are correlated over time and invalidate these shares.

Second, weak instruments might exacerbate the problem of inconsistent estimates when using the two selected shares. Northern European and Baltic migration accounted for a small fraction of migration, as compared to the large migrant groups, such as Mexicans or Indians. Therefore, trying to predict more recent \textit{overall} migration, where their fraction is even less empirically relevant, as is the case especially for Scandinavian migration must result in a low correlation and therefore in weak instruments.

\citet{Goldsmith-Pinkham2020Bartik} show sensitivity-to-misspecification weights that illustrate how the overall bias changes as a certain share's invalidity increases. The shares we identify as being invalid are not among the top-5 sensitivity-to-misspecification weights in their migration example. This shows how small and unsuspicious shares might lead to misleading results and how our method can help in identifying them. 

\section{Conclusion}\label{sec:Conclusion}

We have proposed a novel method to select valid instruments. This method is particularly helpful in cases where the number of candidate instruments is large and tests of overidentifying restrictions reject. 
The method was demonstrated for estimating the effects of immigration on wages but it could be useful for many additional applications, such as estimating the returns to education, in environmental economics and in the study of the effects of health exposures on outcomes with the help of genetic instruments.\footnote{This field of epidemiology is called Mendelian Randomization \citep{Davey2014Mendelian}.}

The strengths of our method are that it extends straightforwardly to the multiple endogenous regressor case and even without a pre-selection step it performs well in settings with weak IVs, avoiding pre-test bias. 
It is still an open question how to deal with local average treatment effects (LATEs). If LATEs are clustered in different groups and all IVs are valid, AHC can be shown to find the set of different LATEs consistently. Another setting where AHC can retrieve an interesting estimate is when the largest group consists of valid IVs and the others of a mixture of LATEs with invalidity. An important contribution would be to disentangle violations of the LATE assumptions and heterogeneous effects in general cases. 
To improve the method one could also account for the variance of each just-identified estimator in the selection algorithm, and apply it in nonlinear models. We leave these questions for future research. 

\bigskip
\begin{center}
{\large\bf SUPPLEMENTARY MATERIAL}
\end{center}

\begin{description}

\item[Online Appendix:]
\begin{itemize}
\item Details on the method
\item Proofs of all Lemmas and Theorems
\item Additional simulations for one regressor case, and for multiple regressors
\end{itemize}

\item[R-code for  AHC:] R-Code to perform the methods. (R-file, will be made available on GitHub)

\end{description}

\begin{center}
	{\large\bf CONFLICT OF INTEREST STATEMENT}
\end{center}

We, the authors, confirm that we have no affiliations with or involvement in any organization or entity with any financial interest (such as honoraria; educational grants; participation in speakers’ bureaus; membership, employment, consultancies, stock ownership, or other equity interest; and expert testimony or patent-licensing arrangements), or non-financial interest (such as personal or professional relationships, affiliations, knowledge or beliefs) in the subject matter or materials discussed in this manuscript.

\newpage
\printbibliography

	\begin{appendices}
	
	\section{Extensions and Discussion}\label{sec:Extensions}
In this Appendix, we provide the theoretical result that the algorithm covers the true model (A.1) and discuss its performance in presence of local violations (\ref{app:localviolations}) and weak instruments (\ref{sec:Weak}). Discussions of different proximity measures and computational complexity can be found in Appendices \ref{sec:ProximityMeasures} and \ref{sec:complexity} respectively.

\subsection{Selection path covers true model}\label{sec:multiple}

To better understand the results of theorems \ref{th:ConsistentSelection} and \ref{th:AsymptoticOracleDistribution}, we introduce and discuss a Lemma. Suppose Algorithm \ref{algo:ward} decides how to merge two of the three clusters $\mathcal{S}_j$, $\mathcal{S}_k$ and $\mathcal{S}_l$, where all the IV combinations associated with the just-identified estimators in $\mathcal{S}_j$ and $\mathcal{S}_k$ are from the same family $\mathcal{F}_{q}$. For $\mathcal{S}_l$, however, it contains at least one estimator such that the corresponding IV combination is from a family other than $\mathcal{F}_q$. The following Lemma establishes that asymptotically, Algorithm \ref{algo:ward} merges $\mathcal{S}_j$ and $\mathcal{S}_k$.

\begin{lemma}\label{lemma:AssignToFamily1}
Let $\mathcal{S}_j$ and $\mathcal{S}_k$ be two clusters such that any just-identified estimator $\hat{\bm{\beta}}_{[j]}$ that is contained in $\mathcal{S}_j$ and $\mathcal{S}_k$ satisfies $[j] \in \mathcal{F}_q$. Let $\mathcal{S}_l$ be a cluster such that $\exists \hat{\bm{\beta}}_{[l]}: \hat{\bm{\beta}}_{[l]} \in \mathcal{S}_l$ and $[l] \in \mathcal{F}_r$ with $\mathbf{r} \neq \mathbf{q}$. Under assumptions \ref{ass:RankAssumption}, \ref{ass:FirstStageSingle}, \ref{ass:ErrorStructure}, \ref{ass:Asymptotics}, \ref{ass:ZwNormal}, \ref{ass:familyplurality} in Algorithm \ref{algo:ward}, if merging two of $\mathcal{S}_j$, $\mathcal{S}_k$ and $\mathcal{S}_l$, then $\mathcal{S}_j$ and $\mathcal{S}_k$ are merged with probability approaching 1.
\end{lemma}
\noindent 
In Algorithm \ref{algo:ward}, we start from the number of clusters $K = \binom{J}{P}$. For each step onward, according to Step 3 in Algorithm \ref{algo:ward}, there would be two clusters joining with each other and forming a new cluster. 

Based on Lemma \ref{lemma:AssignToFamily1}, along the path of Algorithm \ref{algo:ward}, members of different families will not be joined with each other until all the members from the same family have been merged into one cluster. If for each family, all the just-identified estimators associated with the IV combinations in the family have been merged into the same cluster, then we know that the total number of clusters is $K = Q$. This implies that when the number of clusters is smaller than $Q$, at least one cluster contains estimators that use IV-combinations from different families. If the number of clusters is larger than $Q$, then the estimated clusters are subsets of families. 

\begin{corollary} \label{coro:MergingClustersInOwnFamily}
Under assumptions \ref{ass:RankAssumption} to \ref{ass:familyplurality}, in steps 3 and 4 of Algorithm \ref{algo:ward}: 
\begin{equation*}\label{eq:lemma}
\text{When }  \binom{J}{P} \geq K \geq Q, \quad \forall k = 1, ..., K:  \quad  lim \, P(\hat{\mathcal{F}}_{k} \subseteq \mathcal{F}_q) = 1
\end{equation*}
\end{corollary}
\noindent where family $\hat{\mathcal{F}}_k$ corresponds to cluster $\mathcal{S}_k$.

\noindent 
To better understand why this is the case, consider the following analogy. 
There are $N$ guests ($\binom{J}{P}$ just-identified estimates) which belong to $Q$ families. These $N$ people live in a hotel, which has $N$ rooms (clusters). Each day, one room disappears, and one of the people needs to move into the room of some other guest. The people in a family have closer ties, so the person whose room disappears will move into the room of somebody from their own family. This goes on until each family is living respectively in one crowded room. The hotel now continues to shrink. Only now people from different families are merged together into the same rooms. The largest family can be detected, before people from different families start to merge. 

In Algorithm \ref{algo:ward}, the number of clusters starts with $K = \binom{J}{P}$ and ends with $K = 1$. For each step in between, the number of clusters decreases by 1, hence there must be a step where $K = Q$. Based on Lemma \ref{lemma:AssignToFamily1} and Corollary \ref{coro:MergingClustersInOwnFamily}, estimators from different families are joined together only when all elements of their own family have been completely joined to their clusters. This implies that in particular when $K=Q$, there would be a cluster such that all the just-identified estimators in this cluster are estimated by all the valid instruments. Therefore, the path generated by Algorithm \ref{algo:ward} contains the true family with probability going to 1 as there must be one step such that $K=Q$.
\begin{corollary}\label{coro:GroupStructureRetrieved}
When $K = Q$, $lim \, P(\hat{\mathcal{F}}_k = \mathcal{F}_q) = 1 \quad \forall k, q$.
\end{corollary}

\noindent 
The theoretical results above establish that the selection path generated by Algorithm \ref{algo:ward} covers the family which uses only valid IVs, $\mathcal{F}_0$. In Appendix \ref{app:oracle} we show that by Algorithm \ref{algo:Sargan}, we can locate this $\mathcal{F}_0$ and select the valid instruments consistently. This consistent selection property is summarized in Theorem \ref{th:ConsistentSelection} which holds for $P \geq 1$ under Assumption \ref{ass:RankAssumption} to Assumption \ref{ass:plurality} (\ref{ass:familyplurality}). These assumptions also must hold for Theorem \ref{th:AsymptoticOracleDistribution} to hold.

\subsection{Local violations}\label{app:localviolations}

We provide details for our discussion of simulations with local violations in section \ref{sec:simlocalviolations}. 
For $\kappa=1/2$ without global invalidity ($\alpha=0$) \citet*{Andrews2017Measuring} show that the 2SLS-estimator, $\hat{\beta}_{2SLS}$, is asymptotically biased: 
$$\sqrt{n}(\hat{\beta}_{2SLS} - \beta) \overset{d}{\rightarrow} N(0, \mathbf{V}) + ({\mathbf{Q}_{Zd}}^{\prime}\mathbf{Q}_{ZZ}\mathbf{Q}_{Zd})^{-1} {\mathbf{Q}_{Zd}}^{\prime}\bm{\tau}$$
where $\mathbf{Z'd} \overset{P}{\rightarrow} \mathbf{Q}_{Zd}$ and $\mathbf{V}$ is a variance-covariance matrix. 

With strong violations, the bias is exacerbated. 
Moreover, given that we use the Sargan test in the downward testing procedure, we inspect its behaviour in the model with near exclusion to get a better idea of how our procedure will work. 
Table \ref{tab:Summary-Near} offers a summary of results detailed in propositions in Appendix \ref{app:sarganlocal}. The first column shows the size of local violations $\bm{\tau}$, the second column shows the behavior of the Sargan test when all IVs are globally valid and the third column shows the order of growth, $\delta_\xi$ in $\xi_n = o(n^{\delta_\xi})$, needed for the critical values of the Sargan test, $\xi_{n}$, in order for the test to accept. In columns 4 and 5 we repeat the last two columns for a setting where some IVs are globally valid and others are globally invalid but local invalidity affects all. The order of the critical values denotes how large they need to be in order for the test to reject. With the help of these results, we can get a sense under which circumstances the downward testing procedure continues to work correctly. 

In brief, with instruments from the same group, as long as $\bm{\tau} \leq \frac{\mathbf{c}}{\sqrt{n}}$ and $\xi_n \rightarrow \infty$ the Sargan test accepts as $n \rightarrow \infty$. For mixtures, the statistic will always be $O_p(n)$ and as long as $\xi_{n}=o(n)$, the test rejects asymptotically. For strong violations, i.e. $\bm{\tau} > \frac{\mathbf{c}}{\sqrt{n}}$ we want to reject even with groups of globally valid IVs and hence the critical values should not be too large in order to reject: we should have that ${\xi_{n}} < o(n^{1-2\kappa})$. One special case is given when the degree of violation for globally invalid IVs is strong but it is minor for the globally valid IVs, relative to the rate of growth, $\delta$, of the critical value ($\xi_{n} = o(n^{\delta})$): $\kappa_I \leq \frac{1 - \delta}{2} < \kappa_V$. In this case, the Sargan test rejects for strong violations, but not for sets with minor or no violations of exclusion. It could even be that plurality is violated because the largest group of IVs is globally invalid and the downward testing procedure correctly proceeds to a smaller group of valid IVs which is globally valid with minor violations - an especially advantageous case.
	\begin{table}[!t]
		\caption{Summary of Sargan results}\label{tab:Summary-Near}
		\centering	
		\begin{tabular}{p{1.5cm} cc  cc}
			\toprule
			& \multicolumn{2}{c}{Same group} & \multicolumn{2}{c}{Mixture}\\[5pt]\cmidrule(lr){2-3}\cmidrule(lr){4-5} 
			Local violations & Sargan & Needed to accept & Sargan & Needed to reject \\ [10pt]
			\midrule 
			&&&& \\
			None & $Sar\left(\widehat{\bm{\theta}}_{\mathcal{I}}\right) \overset{d}{\rightarrow}\chi^2_{J-K-1}$ & $\delta_\xi > 0$ & $O_p(n)$ & $\xi_{n} = o(n)$ \\ [10pt]
			Minor & $Sar\left(\widehat{\bm{\theta}}_{\mathcal{I}}\right) \overset{d}{\rightarrow}\chi^2_{J-K-1}$ & $\delta_\xi > 0$ & $O_p(n)$ & $\xi_{n} = o(n)$  \\ [10pt]
			Mild & $Sar\left(\widehat{\bm{\theta}}_{\mathcal{I}}\right) \overset{d}{\rightarrow} O_p(1)$ & $\delta_\xi > 0$ & $O_p(n)$ & $\xi_{n} = o(n)$ \\ [10pt]
			Strong  & $Sar\left(\widehat{\bm{\theta}}_{\mathcal{I}}\right) = O_p(n^{1-2\kappa})$ & ${\xi_{n}} > o(n^{1-2\kappa})$ & $O_p(n)$ & $\xi_{n} = o(n)$ \\  [10pt]
			\bottomrule
		\end{tabular}
		\vspace{0.5cm}
		\begin{minipage}{0.9\textwidth} \small
			\textit{Note:} This table summarizes the behavior of the Sargan test with local violations $\bm{\tau}= \frac{\mathbf{c}}{n^{\kappa}}$. The first column denotes the severity of local violations, depending on $\kappa$. Minor stands for  $ 1/2 <\kappa < \infty$, mild stands for $\kappa=1/2$ and strong stands for $ 0 <\kappa < 1/2$. The second column concerns the behavior of the Sargan test when IVs are in the same group. The third column shows how large the critical values need to be in order for the test to accept. The fourth and fifth columns concern the Sargan test when there is a mixture of IVs from different groups but local invalidity affects all. Properties summarized in this table and their proofs are detailed in section \ref{app:sarganlocal}.
		\end{minipage}
	\end{table}

The AHC method can help complement a number of existing methods that are designed to calculate confidence intervals when there are local violations. 
\citet{Conley2012Plausibly} propose to use possible values of the invalidity vector $\alpha$ to create multiple confidence intervals, then take their union, obtaining a confidence interval with conservative coverage. The main drawback of this is that even with only one strong violation the confidence interval becomes very wide and runs the risk of not being particularly informative. AHC could be used to pre-screen strong violations to make the union of CIs narrower. Similarly, \citet{Kang2022Two} propose the union of CIs obtained estimating overidentified models. The drawback is again that inference can be very conservative. 
As described in the main text, \citet*{Guo2021Post} proposes searching and sampling methods for uniformly valid confidence intervals into which AHC can be incorporated for the single regressor case, but here as well CIs are conservative in practice. 

\subsection{Weak instruments}\label{sec:Weak}
In the simulations in section \ref{sec:weaksimulations} we allow for weak IVs among the candidates, by  defining an instrument $Z_j$ as weak if $\gamma_j = C/\sqrt{n}$ where $C$ is a fixed scalar and $C \neq 0$ as in \citet*{Staiger1997Instrumental}. For consistent selection, we maintain the plurality assumption for \textit{strong and valid} instruments as in \citet{Guo2018Confidence}: the group formed by all the strong and valid instruments is the largest group.\footnote{Note that the largest group now also needs to be strong, while IVs in other groups can be weak. The equivalent holds for the largest family when there are multiple regressors.} Here, we provide some additional heuristic discussion to assist in understanding those results. 

We expect the CIM to select weak invalid instruments as valid, because their CIs are wide. Thus, most of them will be overlapping with all other confidence intervals, and the resulting largest group (the selected set of valid IVs) will contain weak invalid IVs. It is noteworthy that inconsistent selection by CIM can arise especially in settings which seem advantageous at first sight: the valid IVs are strong and the invalid are weak. 

For weak and invalid instruments, it can be shown that their just-identified estimands tend to infinity\footnote{Consider $P = 1$. Let $Z_j$ be a weak and invalid instrument, i.e. $\gamma_j = C/\sqrt{n}$ and $\alpha_j \neq 0$. Following Appendix A.5 in \citet{Windmeijer2021Confidence}, for the just-identified estimator of $Z_j$, denoted by $\hat{\beta}_j$, we have $plim(\hat{\beta}_j) = \beta_j = \beta + \frac{\alpha_j}{\gamma_j} = \beta + plim(\sqrt{n}\frac{\alpha_j}{C})$  with $\alpha_j \neq 0$. Therefore $plim(\hat{\beta}_j) \rightarrow \infty$ as $n \rightarrow \infty$.} and they can be separated from the just-identified estimands of the strong and valid instruments as the latter correspond to the true value of the causal effect. 
This implies that AHC limits the impact of including the selected weak instruments on estimation, because it selects the weak and invalid IVs as invalid. These are the IVs that have the strongest detrimental effect on estimates. If the weak valid instruments are classified as valid, their just-identified estimators are not biased too much from the true value. \citet{Windmeijer2019Two} shows that the 2SLS estimator is a weighted average of all just-identified estimates. The weights for each IV-specific estimate increase with the strength of each IV. 
In this case, the biasing effect of including additional weak valid instruments on the 2SLS estimator would be small as their weights of contribution to the 2SLS estimator are small and their estimates are not far from the strong valid ones. 
In comparison, the HT method uses a first-stage hard thresholding and drops all weak IVs. This forces  the user to select an additional tuning parameter and it is not clear how to choose it optimally. More importantly, this procedure might result in pretesting bias as illustrated in \citet{Andrews2019Weak}. 

Overall, we expect AHC to select invalid instruments as invalid regardless of their strength. The results in section \ref{sec:weaksimulations} illustrate that in Design 3 indeed the weak invalid are selected almost always and the weak valid are selected only 12.2\% of the time, affecting the performance of the estimator but clearly improving over the naive estimator. We leave additional theoretical results on the behaviour of the method with weak IVs for future work.

	\subsection{Different Proximity Measures}\label{sec:ProximityMeasures}
	
	In Algorithm \ref{algo:ward} we have made use of the Euclidean distance to assess the proximity of clusters. One might be worried that the results are sensible to the choice of proximity measure. However, in practice this choice does not seem to play an important role.
	
	Furthermore, Algorithm \ref{algo:ward} computes the weighted Euclidean norm to evaluate the distance between clusters. The choice of linkage and distance definition is associated with a specific choice of the objective function, as discussed in \citet{Ward1963Hierarchical}. The latter aims to minimize the sum of within-cluster variation. In complete linkage, the two most distant elements of two clusters define the distance between the clusters. Alternative ways to assess proximity would be to use the medians or centroids of each cluster. 
	In unreported simulations, available on demand, we considered these variants of the AHC algorithm, using different linkage and distance measures and the results are very similar to those obtained by using the Euclidean distance and the Ward-linkage function. We allow for alternative distance definitions and linkage methods in the R-package we provide.
	
	\subsection{Computational Complexity}\label{sec:complexity}
	Recent implementations of the hierarchical agglomerative clustering algorithm have a computational cost of $O(J^2)$ \citep{Amorim2016Ward}. In the downward testing procedure, a maximum of $J-1$ different models needs to be tested. Therefore, the computational cost of the downward testing algorithm is $O(J^2)$. This is an improvement on the CIM which has a time complexity of $O(J^2log(J))$ and where the maximal number of tests is $J(J-1)/2$.
	
	\doublespacing
	
	\section{Proofs}
	
	\subsection{Properties of just-identified estimates when $P \geq 1$}\label{app:PropertiesOfJustIdentified}
	
	There are $\binom{J}{P}$ just-identified models. We write the corresponding just-identified estimators for $\bm{\beta}$ and $\bm{\alpha}$ analogously to the proof of Proposition A.5 in \citet{Windmeijer2021Confidence} for the case $P = 1$. First, for an arbitrary $[j]$, partition the matrix $\mathbf{Z} = (\mathbf{Z}_1 \quad \mathbf{Z}_2)$, where $\mathbf{Z}_1$ is a $n \times P$ matrix containing the $[j]$-th columns of $\mathbf{Z}$, and $\mathbf{Z}_2$ is a $n \times (J-P)$ matrix containing the remaining columns of $\mathbf{Z}$. $\bm{\gamma} = (\bm{\gamma}_1^\prime \quad \bm{\gamma}_2^\prime)'$ is the equivalent partition of the matrix of first-stage coefficients. $\mathbf{Z^*} = [\hat{\mathbf{D}}\quad \mathbf{Z}_2]$ where $\hat{\mathbf{D}}=\mathbf{P}_Z\mathbf{D}$, then $\mathbf{Z^*} = \mathbf{Z} \hat{\mathbf{H}}$, with
	
	\[
	\mathbf{\hat{H}} =
	\left( {\begin{array}{cc}
			\hat{\bm{\gamma}}_1 & 0 \\
			\hat{\bm{\gamma}}_2 & \mathbf{I}_{J - P}\\
	\end{array} } \right)
	; \quad 
	\mathbf{\hat{H}^{-1}} =
	\left( {\begin{array}{cc}
			\hat{\bm{\gamma}}_1^{-1} & 0 \\
			-\hat{\bm{\gamma}}_2\hat{\bm{\gamma}}_1^{-1} & \mathbf{I}_{J - P}\\
	\end{array} } \right)
	\]
	
	\noindent The just-identified 2SLS estimators using $\mathbf{Z}_{[j]}$ as instruments and controlling for the remaining instruments can be written as 
	\[
	(\hat{\bm{\beta}}_{[j]} \quad \hat{\bm{\alpha}}_{[j]}^\prime)' =  \hat{\mathbf{H}}^{-1}\hat{\bm{\Gamma}} = \hat{\mathbf{H}}^{-1}(\mathbf{Z}' \mathbf{Z})^{-1} \mathbf{Z}' (\mathbf{D}\bm{\beta} + \mathbf{Z} \bm{\alpha} + \mathbf{u}) = \hat{\mathbf{H}}^{-1}(\hat{\bm{\gamma}}\bm{\beta} + \bm{\alpha}+(\mathbf{Z}' \mathbf{Z})^{-1} \mathbf{Z}'\mathbf{u})
	\]
	
	\noindent Note that $\hat{\bm{\gamma}}\bm{\beta}+\bm{\alpha}$ is equal to 
	
	\[
	\left(
	\begin{array}{c}
		\hat{\bm{\gamma}}_1 \bm{\beta} + \bm{\alpha}_1\\
		\hat{\bm{\gamma}}_2 \bm{\beta} + \bm{\alpha}_2\\
	\end{array}
	\right) \text{.}
	\]
	
	\noindent By Assumption \ref{ass:Asymptotics}, we have the following probability limits
	
	\[
	plim(\hat{\bm{\beta}}_{[j]}' \quad \hat{\bm{\alpha}}_{[j]}')'
	=plim (\mathbf{\hat{H}^{-1}} \left(
	\begin{array}{c}
		\hat{\bm{\gamma}}_1 \bm{\beta} + \bm{\alpha}_1\\
		\hat{\bm{\gamma}}_2 \bm{\beta} + \bm{\alpha}_2
	\end{array}
	\right))=\left(
	\begin{array}{c}
		\bm{\beta} + \bm{\gamma}_1^{-1}\bm{\alpha}_1\\
		-\bm{\gamma}_2 \bm{\gamma}_1^{-1}\bm{\alpha}_1 + \bm{\alpha}_2
	\end{array}
	\right)
	\]
	
	\noindent We denote the $\binom{J}{P}$ $P \times 1$-dimensional inconsistency terms as $plim(\hat{\bm{\beta}}_{[j]} - \bm{\beta}) = \bm{\gamma}_{[j]}^{-1}\bm{\alpha}_{[j]} = \mathbf{q}_{[j]}$.
	
	\subsection{$\mathcal{F}_0$ consists of valid IVs only}\label{app:gP}
	
	Next, we show that the family with $\mathbf{q}=\mathbf{0}$ is composed of valid IVs with $\bm{\alpha}_1=\mathbf{0}$, only. Let $\bm{\gamma}$, $\mathbf{Z}$ and $\bm{\alpha}$ be partitioned the same way as in Appendix \ref{app:PropertiesOfJustIdentified}.
	
	\begin{remark}
		$\bm{\alpha}_1 = \mathbf{0}$ is necessary and sufficient for $\mathbf{q} = \mathbf{0}$.
	\end{remark}
	
	\paragraph{Proof:} First prove sufficiency:
	Direct proof: Assume $\bm{\alpha}_1=\mathbf{0}$ holds. $\mathbf{q} = \bm{\gamma}_1^{-1} \bm{\alpha}_1=\mathbf{0}$ follows directly. Second, prove necessity: Proof by contraposition: Assume $\bm{\alpha}_1 \neq \mathbf{0}$, then $\bm{\gamma}_1^{-1}\bm{\alpha}_1 \neq \mathbf{0}$. The latter inequality holds, because 
	otherwise the columns of $\bm{\gamma}_1^{-1}$ would be linearly dependent, $\bm{\gamma}_1^{-1}$ would not invertible and hence $(\bm{\gamma}_1^{-1})^{-1} = \bm{\gamma}_1$ would not exist, which it clearly does, by Assumption 1.a. $\qed$
	
	\noindent 
	This also implies that $\mathcal{F}_0$ consists of valid IVs only and all combinations $[j]: \bm{\gamma}_1^{-1}\bm{\alpha}_1=\mathbf{0}$ are elements of $\mathcal{F}_0$. 
	Hence, the following remark directly follows:
	
	\begin{remark}
		$|\mathcal{F}_0| = \binom{g}{P}$.
	\end{remark}
	
	\subsection{Oracle Properties}\label{app:oracle}
	This section gives proofs for Lemma \ref{lemma:AssignToFamily1} and Theorem \ref{th:ConsistentSelection}. All proofs apply for the general case that $P \geq 1$. 
	\subsubsection{Proof of Lemma 1}\label{app:ProofOfLemma1}
	
	Overall, we want to show that the probability that a cluster $\mathcal{S}_j$ with elements from the true underlying partition $\mathcal{S}_{0q}$ is merged with a cluster with elements from the same partition $\mathcal{S}_{0q}$ goes to 1. 
	The proof is structured as follows:
	
	\begin{enumerate}
		\item We note that the means of clusters which are associated with elements from the same family also converge to the same vector as each estimator in the cluster.
		\item Merging two clusters which are associated only with elements from the same family is equivalent to the two clusters having minimal distance.
		\item We show that clusters which are associated with members from the same family have distance zero and clusters which are associated with elements from different families have non-zero distance, with probability going to one.
	\end{enumerate}
	\begin{proof}
		
		\textit{Part 1}: 
		Consider
		\begin{align*}
			\begin{split}
				[j], [k] \in \mathcal{F}_{q} \, ,\quad \mathbf{q} \in \mathbb{R}^P\\
				[l] \in \mathcal{F}_r \, ,\quad \mathbf{r} \in \mathbb{R}^P, \quad \mathbf{r} \neq \mathbf{q}\\
			\end{split}
		\end{align*}
		Under Assumptions 1 - 5:
		\begin{align}
			\begin{split}
				plim(\hat{\bm{\beta}}_{[j]}) = plim(\hat{\bm{\beta}}_{[k]}) = \mathbf{q}\\
				plim(\hat{\bm{\beta}}_{[l]}) = \mathbf{r}
			\end{split}
		\end{align}
		Let $\mathcal{S}_j$ and $\mathcal{S}_k$ be clusters associated with elements from the same family: $\mathcal{S}_j$, $\mathcal{S}_k \subset \mathcal{S}_{0q}$ and $\mathcal{S}_{l} \subset \mathcal{S}_{0r}$.
		\begin{equation}\label{eq:conv}
			plim \,\, \bar{\mathcal{S}}_j = plim \frac{ \sum\limits_{\hat{\beta}_{[j]} \in \mathcal{S}_{j}} \bm{\hat{\beta}_{[j]}}}{|\mathcal{S}_{j}|} = \frac{|\mathcal{S}_{j}|\mathbf{q}}{|\mathcal{S}_{j}|} 
		\end{equation}
		and hence $$plim(\bar{\mathcal{S}}_{j}) = \mathbf{q}\text{.}$$
		
		\noindent
		\textit{Part 2:}
		Consider the case where the Algorithm decides whether to merge two clusters, $\mathcal{S}_j$ and $\mathcal{S}_k$, containing estimators using combinations from the same family, or to merge two clusters from different underlying partitions, $\mathcal{S}_j$ and $\mathcal{S}_l$. The two clusters which are closest in terms of their weighted squared Euclidean distance are merged first. Hence, we need to consider the distances between  $\mathcal{S}_j$ and $\mathcal{S}_k$,  $\mathcal{S}_j$ and $\mathcal{S}_l$,  as well as $\mathcal{S}_k$ and $\mathcal{S}_l$.
		
		$\mathcal{S}_j$ is merged with a cluster with elements of its own $\mathcal{S}_{0q}$ iff
		$\frac{|\mathcal{S}_j||\mathcal{S}_k|}{|\mathcal{S}_j| + |\mathcal{S}_k|}||\bar{\mathcal{S}}_j - \bar{\mathcal{S}}_k||^2 < \frac{|\mathcal{S}_j||\mathcal{S}_l|}{|\mathcal{S}_j| + |\mathcal{S}_l|}||\bar{\mathcal{S}}_j - \bar{\mathcal{S}}_l||^2$. The following two are hence equivalent
		
		\begin{equation*}
			lim \, P (\mathcal{S}_j \cup \mathcal{S}_k = \mathcal{S}_{jk} \subseteq \mathcal{S}_{0q}) = 1
		\end{equation*}
		\begin{equation}\label{eq:Lemma1ToShow}
			\Leftrightarrow \quad  lim \, P(\frac{|\mathcal{S}_j||\mathcal{S}_k|}{|\mathcal{S}_j| + |\mathcal{S}_k|}||\bar{\mathcal{S}}_j - \bar{\mathcal{S}}_k||^2 < \frac{|\mathcal{S}_j||\mathcal{S}_l|}{|\mathcal{S}_j| + |\mathcal{S}_l|}||\bar{\mathcal{S}}_j - \bar{\mathcal{S}}_l||^2) = 1
		\end{equation}
		where $\mathcal{S}_{jk}$ is the new merged cluster. 
		
		\textit{Part 3}: We want to prove equation \eqref{eq:Lemma1ToShow} in the following. We can then prove $lim \, P(\frac{|\mathcal{S}_j||\mathcal{S}_k|}{|\mathcal{S}_j| + |\mathcal{S}_k|}||\bar{\mathcal{S}}_k - \bar{\mathcal{S}}_j||^2 < \frac{|\mathcal{S}_k||\mathcal{S}_l|}{|\mathcal{S}_k| + |\mathcal{S}_l|}||\bar{\mathcal{S}}_k - \bar{\mathcal{S}}_l||^2) = 1$ by changing the subscripts. 
		First, define $a_n = \frac{|\mathcal{S}_j||\mathcal{S}_k|}{|\mathcal{S}_j| + |\mathcal{S}_k|}||\bar{\mathcal{S}}_j - \bar{\mathcal{S}}_k||^2$, $b_n=\frac{|\mathcal{S}_j||\mathcal{S}_l|}{|\mathcal{S}_j| + |\mathcal{S}_l|}||\bar{\mathcal{S}}_j - \bar{\mathcal{S}}_l||^2$ and $c=\frac{|\mathcal{S}_j||\mathcal{S}_l|}{|\mathcal{S}_j| + |\mathcal{S}_l|} (\mathbf{q}-\mathbf{r})' (\mathbf{q}-\mathbf{r})$. 
		Under \eqref{eq:conv}
		\begin{align*}
			\begin{split}
				plim(a_n)& =  \bm{0}\\
				plim(b_n)& = c\\
			\end{split}
		\end{align*}
		
		To show: $\underset{n \rightarrow \infty}{\lim}\, P (a_n<b_n)=1$. 
		Since $plim(a_n) - plim(b_n) = -c$. By the definition of convergence in probability, we have $\lim_{n \rightarrow \infty} P (|a_n - b_n + c| < \epsilon) = 1$ 
		holds for any $\epsilon > 0$. Let $\epsilon  = c/2$, then we have $\lim_{n \rightarrow \infty} P (-c/2 <a_n - b_n + c < c/2) = 1$, $\lim_{n \rightarrow \infty} P (-3c/2 <a_n - b_n < -c/2 < 0) = 1$ and therefore, 
		$\lim_{n \rightarrow \infty} P (a_n - b_n < 0) = 1$.				\end{proof}
	
	\subsubsection{Proof of Theorem \ref{th:ConsistentSelection}}
	\begin{proof}
		The proof of Theorem \ref{th:ConsistentSelection} is structured as follows:
		\begin{enumerate}
			\item We show that asymptotically the selection path generated by Algorithm \ref{algo:ward} contains $\mathcal{F}_0$, the family formed by all the valid instrumental variables.
			\item We show that Algorithm \ref{algo:Sargan} can recover $\mathcal{F}_0$ from the selection path from Algorithm \ref{algo:ward}.
		\end{enumerate}
		\textit{Part 1} follows from Corollary \ref{coro:GroupStructureRetrieved} directly.
		
		\noindent \textit{Part 2}: Firstly, we establish the properties of the Sargan statistic.
		The following two equations can be also found in \citet[][p.10]{Windmeijer2021Confidence}. Let $\mathcal{I}$ be the true set of invalid instruments and $\mathcal{V}$ be the true set of valid instruments. The oracle model is 
		
		\begin{equation*}
			\mathbf{y} = \mathbf{D}\bm{\beta} + \mathbf{Z}_{\mathcal{I}} \bm{\alpha}_{\mathcal{I}} + \mathbf{u} = \mathbf{X}_\mathcal{I} \bm{\theta}_\mathcal{I} + \mathbf{u}
		\end{equation*}
		with $\mathbf{X}_\mathcal{I} = \left[\mathbf{D} \quad \mathbf{Z}_{\mathcal{I}}\right]$ and $\bm{\theta}_{\mathcal{I}} = \left[\bm{\beta} \quad \bm{\alpha}_{\mathcal{I}}'\right]^\prime$, the Sargan test statistic is given by 
		
		\begin{equation}
			S(\hat{\bm{\theta}}_{\mathcal{I}})=\frac{\hat{\mathbf{u}}(\hat{\bm{\theta}}_{\mathcal{I}})^\prime\mathbf{Z}_{\mathcal{I}}(\mathbf{Z}_{\mathcal{I}}^\prime\mathbf{Z}_{\mathcal{I}})^{-1}\mathbf{Z}_{\mathcal{I}}^\prime\hat{\mathbf{u}}(\hat{\bm{\theta}}_{\mathcal{I}})}{\hat{\mathbf{u}}(\hat{\theta}_{\mathcal{I}})^\prime\hat{\mathbf{u}}(\hat{\bm{\theta}}_{\mathcal{I}})/n}
		\end{equation}
		where $\hat{\mathbf{u}}(\hat{\bm{\theta}})=\mathbf{y} - \mathbf{X}_{\mathcal{I}}\hat{\bm{\theta}}_{\mathcal{I}}$, with $\hat{\bm{\theta}}_{\mathcal{I}}$ the 2SLS estimator of $\bm{\theta}_{\mathcal{I}}$.
		
		\noindent Let $\hat{\mathcal{I}}$ be the estimated set of invalid instruments and $\hat{\mathcal{V}}$ be the estimated set of valid instruments where $\hat{\mathcal{I}} = \mathcal{J} \setminus \hat{\mathcal{V}}$. Following \citet{Windmeijer2021Confidence}, the Sargan statistic has the following properties, where $\mathcal{V}^s$ is a set of valid IVs with $|\mathcal{V}^s| \leq |\mathcal{V}|$: 
		
		\begin{property}Properties of the Sargan statistic\label{prop:Sargan}
			\begin{enumerate}
				\item For all the $|\mathcal{\hat{\mathcal{V}}}| \choose P$ combinations of the instruments from $\hat{\mathcal{V}}$, if the IVs contained in them belong to the same family, then: $S(\hat{\bm{\theta}}_\mathcal{\hat{\mathcal{I}}}) \overset{d}{\rightarrow} \chi^2_{|\mathcal{V}^s| - P}$
				\item For all the $|\mathcal{\hat{\mathcal{V}}}| \choose P$ combinations of the instruments from $\hat{\mathcal{V}}$, if the IVs contained in them belong to a mixture of families, then: $S(\hat{\bm{\theta}}_\mathcal{\hat{\mathcal{I}}})=O_p(n)$.
			\end{enumerate}
		\end{property}
		
		\noindent 
		With these properties we can show that the downward testing procedure described in Algorithm \ref{algo:Sargan} selects the valid instruments consistently with the critical values $\xi_{n, |\mathcal{V}^s| - P} \rightarrow \infty \text{ for } n \rightarrow \infty \text{, and } \xi_{n, |\mathcal{V}^s| - P}=o(n)\text{.}$ Let the number of clusters formed in Algorithm \ref{algo:ward} at some certain step be $K$, e.g. at Step 1, $K = {J \choose P}$ and at Step 2, $K = {J \choose P}-1$ etc. Let the true number of families be $Q$. Consider applying the Sargan test to the model selected by the largest cluster at each step under the following scenarios:
		\begin{enumerate}
			\item $1 \leq K < Q$. For each of these steps, the largest cluster is either associated with a mixture of different families, or with one family.
			\begin{itemize}
				\item Consider the case where the largest cluster is associated with a mixture of different families. Then by Property \ref{prop:Sargan} and $\xi_{n, |\mathcal{V}^s| - P}=o(n)$, we have 
				\begin{equation*}
					\underset{n \rightarrow \infty}{lim} P (S(\hat{\bm{\theta}}_{\hat{\mathcal{I}}}) < \xi_{n, |\mathcal{V}^s| - P}) = 0 \text{.}
				\end{equation*}
				In this case, asymptotically the Sargan test would be rejected and the downward testing procedure moves to the next step.
				\item Consider the case where the largest cluster is associated with one family. Then this family must be $\mathcal{F}_0$ as by Assumption \ref{ass:familyplurality}, $\mathcal{F}_0$ is the largest family among all $Q$ families. Then following Property \ref{prop:Sargan} and $\xi_{n, |\mathcal{V}^s| - P} \rightarrow \infty$ for the Sargan test we have
				\begin{equation}\label{eq:sarganpass}
					\underset{n \rightarrow \infty}{lim} P (S(\hat{\bm{\theta}}_{\hat{\mathcal{I}}}) < \xi_{n, |\mathcal{V}^s| - P}) = 1 \text{.}
				\end{equation}
				indicating that $\mathcal{V}$ would be selected as the set of valid instruments asymptotically.
			\end{itemize}
			\item $K = Q$. By Corollary \ref{coro:GroupStructureRetrieved} we know that the $K$ clusters are associated with the $Q$ families respectively, and by Assumption \ref{ass:familyplurality}, the cluster associated with $\mathcal{F}_0$ is the largest cluster. Then applying the Sargan test at this step would be testing all the valid instruments, hence we also have Equation \eqref{eq:sarganpass} and Algorithm \ref{algo:Sargan} selects $\mathcal{V}$ as the set of valid instruments.
		\end{enumerate}
		To summarize, asymptotically, at steps $1 \leq K < Q$, Algorithm \ref{algo:Sargan} only stops when $\mathcal{F}_0$ forms the largest cluster and hence selects the oracle model, otherwise it moves to step $K = Q$ and selects the oracle model.
		\noindent 
		Combine \textit{Part 1} and \textit{Part 2}, we prove Theorem \ref{th:ConsistentSelection}.
		
	\end{proof}
	
	\subsection{Sargan statistic under local violations}\label{app:sarganlocal}
	In this appendix, we provide detailed statements and proofs for the results discussed in section \ref{app:localviolations}.	Note that here we refer to model \ref{eq:StructuralLocalViolations} and that the Sargan test statistic is
	\begin{equation*}
		Sar\left(\widehat{\bm{\theta}}_{\mathcal{I}}\right)=\frac{\widehat{\mathbf{u}}(\widehat{\bm{\theta}}_{\mathcal{I}})^{\prime}\mathbf{Z}\left(\mathbf{Z}^{\prime}\mathbf{Z}\right)^{-1}\mathbf{Z}^{\prime}\widehat{\mathbf{u}}(\widehat{\bm{\theta}}_{\mathcal{I}})}{\widehat{\mathbf{u}}(\widehat{\bm{\theta}}_{\mathcal{I}})^{\prime}\widehat{\mathbf{u}}(\widehat{\bm{\theta}}_{\mathcal{I}})/n}.
	\end{equation*}
	\begin{proposition}\label{prop:Sargan-GlobValidLocValid}
		If $\bm{\tau}=0$, then 
		\begin{equation}
			Sar\left(\widehat{\bm{\theta}}_{\mathcal{I}}\right) \overset{d}{\rightarrow} \bm{\zeta}^{\prime} \mathbf{A} \bm{\zeta} \sim \chi^2_{J-|\mathcal{I}|-1}
		\end{equation}
	\end{proposition}
	\noindent where $\bm{\zeta} \sim N(0, \mathbf{I})$ and $\mathbf{A}$ is a finite square matrix. The proof follows by standard arguments but we still reproduce it here so as to let the reader follow the other propositions more easily.
	
	\paragraph{Proof of Proposition \ref{prop:Sargan-GlobValidLocValid}}
	We know that 
	$\frac{1}{n} \mathbf{Z}^{\prime}\mathbf{u} \overset{P}{\rightarrow} 0$, $\frac{1}{n} \mathbf{Z}^{\prime}\mathbf{Z} \overset{P}{\rightarrow} \mathbf{Q}_{ZZ}$, $\frac{1}{n} \mathbf{Z}^{\prime}\mathbf{X} \overset{P}{\rightarrow} \mathbf{Q}_{ZX}$ and $\frac{1}{\sqrt{n}} \mathbf{Z}^{\prime}\mathbf{u} \overset{d}{\rightarrow} N(0, \sigma_{u}^2 \mathbf{Q}_{ZZ})$, as $n \rightarrow \infty$. 	
	Note that if all invalid instruments are selected correctly and controlled for, the estimator of $\bm{\theta}$ is consistent:
	
	\begin{equation}\label{eq:ConsistencyOfTheta-Near}
		\widehat{\bm{\theta}}_{\mathcal{I}} \overset{P}{\rightarrow} \bm{\theta}_{\mathcal{I}} \text{.}
	\end{equation}
	
	\noindent The residual can be written as: 
	\begin{align}\label{eq:ResidAllValid}
		\begin{split}
			\widehat{\mathbf{u}}(\widehat{\bm{\theta}}_{\mathcal{I}}) & = \mathbf{y} - \mathbf{X}_{\mathcal{I}}\widehat{\bm{\theta}} = \mathbf{X}_{\mathcal{I}}\bm{\theta}_{\mathcal{I}} + \mathbf{u} - \mathbf{X}_{\mathcal{I}}\widehat{\bm{\theta}}\\
			& = \mathbf{X}_{\mathcal{I}}\bm{\theta}_{\mathcal{I}} + \mathbf{u} - \mathbf{X}_{\mathcal{I}}({\mathbf{X}_{\mathcal{I}}}^{\prime} \mathbf{P}_Z\mathbf{X}_{\mathcal{I}})^{-1}{\mathbf{X}_{\mathcal{I}}}^{\prime} \mathbf{P}_Z\mathbf{y}\\ 
			& = \mathbf{X}_{\mathcal{I}}\bm{\theta}_{\mathcal{I}} + \mathbf{u} - \mathbf{X}_{\mathcal{I}}({\mathbf{X}_{\mathcal{I}}}^{\prime} \mathbf{P}_Z\mathbf{X}_{\mathcal{I}})^{-1}{\mathbf{X}_{\mathcal{I}}}^{\prime} \mathbf{P}_Z(\mathbf{X}_{\mathcal{I}}\bm{\theta}_{\mathcal{I}} + \mathbf{u}) \\ 
			& = \mathbf{u} - \mathbf{X}_{\mathcal{I}}({\mathbf{X}_{\mathcal{I}}}^{\prime} \mathbf{P}_Z\mathbf{X}_{\mathcal{I}})^{-1}{\mathbf{X}_{\mathcal{I}}}^{\prime} \mathbf{P}_Z\mathbf{u}\text{.}
		\end{split}
	\end{align}
	Note that in $\mathbf{u}^{\prime}\mathbf{P}_Z\mathbf{X}_{\mathcal{I}} ({\mathbf{X}_{\mathcal{I}}}^{\prime} \mathbf{P}_Z\mathbf{X}_{\mathcal{I}})^{-1} {\mathbf{X}_{\mathcal{I}}}^{\prime}\mathbf{X}_{\mathcal{I}}({\mathbf{X}_{\mathcal{I}}}^{\prime} \mathbf{P}_Z\mathbf{X}_{\mathcal{I}})^{-1}{\mathbf{X}_{\mathcal{I}}}^{\prime} \mathbf{P}_Z\mathbf{u}$ we have that $plim(\mathbf{u}^{\prime}\mathbf{Z}) = 0$, while $plim(\mathbf{Z}^{\prime}\mathbf{Z})$, $plim({\mathbf{X}_{\mathcal{I}}}^{\prime}\mathbf{Z})$ and $plim({\mathbf{X}_{\mathcal{I}}}^{\prime}\mathbf{X}_{\mathcal{I}})$ all converge against finite, non-zero matrices. In this term, we have an uneven number of terms that converge, two of them to zero. Therefore, dividing by $n$ in the limit leads to a product of finite matrices with zero. All terms except for $\mathbf{u}^{\prime}\mathbf{u}/n$ disappear for this reason. 
	Hence
	\begin{equation*}
		\widehat{\mathbf{u}}(\widehat{\bm{\theta}}_{\mathcal{I}})^{\prime}\widehat{\mathbf{u}}(\widehat{\bm{\theta}}_{\mathcal{I}})/n \overset{P}{\rightarrow} \sigma_u^2 \text{.}
	\end{equation*}
	Also, the numerator of the Sargan statistic can be written as
	\begin{equation}
		\widehat{\mathbf{u}}(\widehat{\bm{\theta}}_{\mathcal{I}})'\mathbf{P}_Z\widehat{\mathbf{u}}(\widehat{\bm{\theta}}_{\mathcal{I}}) = 	\widehat{\mathbf{u}}(\widehat{\bm{\theta}}_{\mathcal{I}})'\mathbf{Z} (\mathbf{Z}^{\prime}\mathbf{Z})^{-1/2}(\mathbf{Z}^{\prime}\mathbf{Z})^{-1/2}\mathbf{Z}^{\prime}	\widehat{\mathbf{u}}(\widehat{\bm{\theta}}_{\mathcal{I}}) \text{.}\\
	\end{equation}
	Plugging in \eqref{eq:ResidAllValid}, we get
	\begin{equation}\label{eq:RootOfNumerator}
		(\mathbf{Z}^{\prime}\mathbf{Z})^{-1/2}\mathbf{Z}^{\prime}	\widehat{\mathbf{u}}(\widehat{\bm{\theta}}_{\mathcal{I}}) = (\mathbf{I} - (\mathbf{Z}^{\prime}\mathbf{Z})^{-1/2}\mathbf{Z}^{\prime}\mathbf{X}_\mathcal{I}({\mathbf{X}_\mathcal{I}}^{\prime} \mathbf{P}_Z\mathbf{X}_\mathcal{I})^{-1}{\mathbf{X}_\mathcal{I}}^{\prime} \mathbf{Z} (\mathbf{Z}^{\prime}\mathbf{Z})^{-1/2} )(\mathbf{Z}^{\prime}\mathbf{Z})^{-1/2}\mathbf{Z}^{\prime}\mathbf{u}
	\end{equation}
	With the assumptions shown at the beginning, it can be seen that
	\begin{equation}
		(\mathbf{Z}^{\prime}\mathbf{Z})^{-1/2}\mathbf{Z}^{\prime}\mathbf{u} \overset{d}{\rightarrow} N(0, \sigma^2_u\mathbf{I})
	\end{equation}
	as $(\frac{\mathbf{Z'Z}}{n})^{-1/2}= \mathbf{Q}_{ZZ}^{-1/2}$, $\frac{\mathbf{Z'u}}{\sqrt{n}} \overset{d}{\rightarrow} N(0, \sigma_{u}^2 \mathbf{Q}_{ZZ}) $ and hence
	\begin{equation}
		\frac{(\mathbf{Z}^{\prime}\mathbf{Z})^{-1/2}\mathbf{Z}^{\prime}\mathbf{u}}{\sqrt{\widehat{\mathbf{u}}(\widehat{\bm{\theta}}_{\mathcal{I}})^{\prime}\widehat{\mathbf{u}}(\widehat{\bm{\theta}}_{\mathcal{I}})/n}} \overset{d}{\rightarrow} N(0, \mathbf{I})
	\end{equation}
	Consider $\mathbf{I} - (\mathbf{Z}^{\prime}\mathbf{Z})^{-1/2}\mathbf{Z}^{\prime}\mathbf{X}_{\mathcal{I}}({\mathbf{X}_\mathcal{I}}^{\prime} \mathbf{P}_Z\mathbf{X}_\mathcal{I})^{-1}{\mathbf{X}_\mathcal{I}}^{\prime} \mathbf{Z} (\mathbf{Z}^{\prime}\mathbf{Z})^{-1/2}$, a symmetric, idempotent projection matrix that goes to a finite square matrix $\mathbf{A}$ in probability and hence: 
	\begin{equation}
		Sar\left(\widehat{\bm{\theta}}_{\mathcal{I}}\right) \overset{d}{\rightarrow} \bm{\eta}^{\prime} \mathbf{A} \bm{\eta} \sim \chi^2_{J-|\mathcal{I}|-1}
	\end{equation}
	where $\bm{\eta} \sim N(0, \mathbf{I})$. This holds by Theorem 2 in \citet[][p.57-59]{Searle1971Linear}. $\qed$
	
	\subsubsection{Globally valid, locally invalid}\label{sec:OnlyValid-Violations}
	
	Next, we assume that the globally valid IVs have been selected but there is local invalidity as in the model in equation \ref{eq:StructuralLocalViolations}. 
	\begin{equation*}
		\mathbf{y} = \mathbf{d}\beta + \mathbf{Z}_{\mathcal{I}} \bm{\alpha}_{\mathcal{I}} + \mathbf{Z}\bm{\tau}  + \mathbf{u} = \mathbf{X}_{\mathcal{I}} \bm{\theta}_{\mathcal{I}} + \mathbf{Z}\bm{\tau} + \mathbf{u}
	\end{equation*}
	
	\begin{proposition}\label{prop:Sargan-GlobValidLocInvalid}
		If $\bm{\tau} = o_p(n^{-1/2})$, then 
		\begin{equation*}
			Sar\left(\widehat{\bm{\theta}}_{\mathcal{I}}\right) \overset{d}{\rightarrow} \chi^2_{J - |\mathcal{I}| - 1} \text{ .}
		\end{equation*}
	\end{proposition}
	
	\paragraph{Proof of Proposition \ref{prop:Sargan-GlobValidLocInvalid}}
	\noindent The 2SLS estimator is
	\begin{align}\label{eq:ConsistencyOfTheta-LocalViolations}
		\begin{split}
			\widehat{\bm{\theta}}_{\mathcal{I}} & = ({\mathbf{X}_{\mathcal{I}}}^{\prime} \mathbf{P}_Z \mathbf{X}_{\mathcal{I}})^{-1} {\mathbf{X}_{\mathcal{I}}}^{\prime} \mathbf{P}_Z \mathbf{y} \\ 
			&= ({\mathbf{X}_{\mathcal{I}}}^{\prime} \mathbf{P}_Z \mathbf{X_{\mathcal{I}}})^{-1} {\mathbf{X_{\mathcal{I}}}}^{\prime} \mathbf{P}_Z (\mathbf{X}_{\mathcal{I}}\bm{\theta}_{\mathcal{I}} + \mathbf{Z}\bm{\tau} + \mathbf{u})\\
			& = \bm{\theta}_{\mathcal{I}} + (\mathbf{X_{\mathcal{I}}}^{\prime} \mathbf{P}_Z \mathbf{X_{\mathcal{I}}})^{-1} \mathbf{X_{\mathcal{I}}}^{\prime} \mathbf{P}_Z (\mathbf{Z}\bm{\tau} + \mathbf{u}) \\ 
			& =  \bm{\theta}_{\mathcal{I}} + ({\mathbf{X}_{\mathcal{I}}}^{\prime} \mathbf{P}_Z \mathbf{X}_{\mathcal{I}})^{-1} {\mathbf{X}_{\mathcal{I}}}^{\prime} \mathbf{Z}\bm{\tau} + ({\mathbf{X}_{\mathcal{I}}}^{\prime} \mathbf{P}_Z \mathbf{X}_{\mathcal{I}})^{-1} {\mathbf{X}_{\mathcal{I}}}^{\prime} \mathbf{P}_Z \mathbf{u}
		\end{split}
	\end{align}
	
	\noindent If $\bm{\tau} = o_p(1)$, then $\widehat{\bm{\theta}}_{\mathcal{I}}\overset{P}{\rightarrow} \bm{\theta}_\mathcal{I}$ and the residual becomes
	
	\begin{align}
		\begin{split}
			\widehat{\mathbf{u}}(\widehat{\bm{\theta}}_{\mathcal{I}}) & = \mathbf{X}_{\mathcal{I}} \bm{\theta}_{\mathcal{I}} + \mathbf{Z}\bm{\tau} + \mathbf{u} - \mathbf{X_{\mathcal{I}}}\widehat{\bm{\theta}}_{\mathcal{I}} \\
			& = \mathbf{Z}\bm{\tau}+ \mathbf{u} - \mathbf{X_{\mathcal{I}}}(\bm{\theta}_{\mathcal{I}} - \widehat{\bm{\theta}}_{\mathcal{I}} )\\  
			& = \mathbf{Z}\bm{\tau}+ \mathbf{u} - \mathbf{X}_{\mathcal{I}}({\mathbf{X}_{\mathcal{I}}}^{\prime} \mathbf{P}_Z\mathbf{X}_{\mathcal{I}})^{-1}{\mathbf{X}_{\mathcal{I}}}^{\prime} \mathbf{Z}\bm{\tau} - \mathbf{X}_{\mathcal{I}}({\mathbf{X}_{\mathcal{I}}}^{\prime} \mathbf{P}_Z\mathbf{X}_{\mathcal{I}})^{-1}{\mathbf{X}_{\mathcal{I}}}^{\prime} \mathbf{P}_Z\mathbf{u}
		\end{split}
	\end{align}
	Hence:
	\begin{align}\label{eq:AllValid-ErrorVariance}
		\begin{split}
			\widehat{\mathbf{u}}(\widehat{\bm{\theta}}_{\mathcal{I}})^{\prime}\widehat{\mathbf{u}}(\widehat{\bm{\theta}}_{\mathcal{I}})/n  & = 
			(\mathbf{Z}\bm{\tau}+ \mathbf{u} - \mathbf{X}_{\mathcal{I}}({\mathbf{X}_{\mathcal{I}}}^{\prime} \mathbf{P}_Z\mathbf{X}_{\mathcal{I}})^{-1}{\mathbf{X}_{\mathcal{I}}}^{\prime} \mathbf{Z}\bm{\tau} - \mathbf{X}_{\mathcal{I}}({\mathbf{X}_{\mathcal{I}}}^{\prime} \mathbf{P}_Z\mathbf{X}_{\mathcal{I}})^{-1}{\mathbf{X}_{\mathcal{I}}}^{\prime} \mathbf{P}_Z\mathbf{u})' \\
			& \quad (\mathbf{Z}\bm{\tau}+ \mathbf{u} - \mathbf{X}_{\mathcal{I}}({\mathbf{X}_{\mathcal{I}}}^{\prime} \mathbf{P}_Z\mathbf{X}_{\mathcal{I}})^{-1}{\mathbf{X}_{\mathcal{I}}}^{\prime} \mathbf{Z}\bm{\tau} - \mathbf{X}_{\mathcal{I}}({\mathbf{X}_{\mathcal{I}}}^{\prime} \mathbf{P}_Z\mathbf{X}_{\mathcal{I}})^{-1}{\mathbf{X}_{\mathcal{I}}}^{\prime} \mathbf{P}_Z\mathbf{u})\\
			& = \frac{\bm{\tau}^{\prime} \mathbf{Z}^{\prime} \mathbf{Z} \bm{\tau}}{n} + \frac{\mathbf{u}^{\prime}\mathbf{u}}{n} - 2\frac{\mathbf{u}^{\prime}\mathbf{X}_{\mathcal{I}}({\mathbf{X}_{\mathcal{I}}}^{\prime}\mathbf{P}_Z \mathbf{\mathbf{X}_{\mathcal{I}})^{-1}{\mathbf{X}_{\mathcal{I}}}^{\prime}\mathbf{Z}} \bm{\tau}}{n} \\
			& \quad - 2\frac{\bm{\tau}^{\prime}\mathbf{Z}^{\prime}\mathbf{X}_{\mathcal{I}}({\mathbf{X}_{\mathcal{I}}}^{\prime}\mathbf{P}_Z \mathbf{\mathbf{X}_{\mathcal{I}})^{-1}}\mathbf{\mathbf{X}_{\mathcal{I}}}^{\prime} \mathbf{Z} \bm{\tau}}{n} \\ 
			& \quad + \frac{\bm{\tau}^{\prime}\mathbf{Z}^{\prime}\mathbf{X}_{\mathcal{I}}({\mathbf{X}_{\mathcal{I}}}^{\prime}\mathbf{P}_Z \mathbf{\mathbf{X}_{\mathcal{I}})^{-1}}\mathbf{\mathbf{X}_{\mathcal{I}}}^{\prime} \mathbf{X}_{\mathcal{I}}({\mathbf{X}_{\mathcal{I}}}^{\prime}\mathbf{P}_Z \mathbf{\mathbf{X}_{\mathcal{I}})^{-1}{\mathbf{X}_{\mathcal{I}}}^{\prime}\mathbf{Z}}\bm{\tau}}{n} \\
			& \quad + o_p(1)\text{.}
		\end{split}
	\end{align}
	The terms that involve ($\frac{\mathbf{u}^{\prime}\mathbf{Z}}{n} \overset{P}{\rightarrow} 0$) go to zero in probability and are subsumed by $o_p(1)$.
	This term goes to $\sigma_u^2$ in probability, when $\bm{\tau} = o_p(1)$. 
	The root of the numerator of the Sargan test, the term from equation \ref{eq:RootOfNumerator}, now becomes
	\begin{align}\label{eq:RewriteEmpiricalMoments-GlobalValidLocalViolations}
		\begin{split}
			(\mathbf{Z}^{\prime}\mathbf{Z})^{-1/2}\mathbf{Z}^{\prime}	\widehat{\mathbf{u}}(\widehat{\bm{\theta}}_{\mathcal{I}}) &= (\mathbf{Z}^{\prime}\mathbf{Z})^{-1/2}\mathbf{Z}^{\prime}(\mathbf{Z}\bm{\tau} + \mathbf{u} - \mathbf{X}_{\mathcal{I}}({\mathbf{X}_{\mathcal{I}}}^{\prime} \mathbf{P}_Z\mathbf{X}_{\mathcal{I}})^{-1}{\mathbf{X}_{\mathcal{I}}}^{\prime} \mathbf{P}_Z(\mathbf{Z}\bm{\tau} + \mathbf{u}) )\\
			& = (\frac{\mathbf{Z}^{\prime}\mathbf{Z}}{n})^{-1/2}\frac{\mathbf{Z}^{\prime}\mathbf{Z}}{\sqrt{n}}\bm{\tau} + (\mathbf{Z}^{\prime}\mathbf{Z})^{-1/2}\mathbf{Z}^{\prime}\mathbf{u} \\ 
			& \quad - (\mathbf{Z}^{\prime}\mathbf{Z})^{-1/2}\mathbf{Z}^{\prime}\mathbf{X}_\mathcal{I}({\mathbf{X}_\mathcal{I}}^{\prime} \mathbf{P}_Z{\mathbf{X}_\mathcal{I}})^{-1}{\mathbf{X}_\mathcal{I}}^{\prime}\mathbf{P}_Z (\mathbf{Z} \bm{\tau} + \mathbf{u}) \text{.}
		\end{split}
	\end{align}
	In order for this term to converge to the same normal distribution as before all summands involving $\bm{\tau}$ have to disappear. This is the case when there are minor violations, i.e. the local violation vector is $\bm{\tau} = o_p(n^{-1/2})$. If that is true, it still holds that
	
	\begin{equation*}
		Sar\left(\widehat{\bm{\theta}}_{\mathcal{I}}\right) \overset{d}{\rightarrow} \chi^2_{J - |\mathcal{I}| - 1} \text{\qed}
	\end{equation*}
	
	\paragraph*{Mild violations: $\bm{\kappa=1/2}$}
	Next, we look at the border case of mild violations, when $\bm{\tau} = \frac{\mathbf{c}}{\sqrt{n}}$. This case was the object of a lot of attention in the nearly exogenous instruments literature, because exogeneity error and sampling error both play a role contemporaneously for the asymptotic distribution. 
	\begin{proposition}\label{prop:Sargan-GlobValidLocInvalidK=0.5}
		If  $\bm{\tau} = \frac{\mathbf{c}}{\sqrt{n}}$, then 
		\begin{equation}\label{eq:Sargan-BorderCase-GlobalValidLocalViolations}
			Sar\left(\widehat{\bm{\theta}}_{\mathcal{I}}\right) \overset{d}{\rightarrow} \chi^2_{J - |\mathcal{I}| - 1}\left(\frac{1}{\sigma_{u}^2} \mathbf{c}^{\prime}{\mathbf{Q}_{ZZ}^{1/2}}^{\prime}\mathbf{A} \mathbf{Q}_{ZZ}^{1/2}\mathbf{c}\right) \text{ .}
		\end{equation}
	\end{proposition}
	\noindent This result corresponds to the ones in \citet[][Theorem 2.1]{Newey1985Generalized} and \citet[][Theorem 2]{Hayakawa2014Alternative} who look at general violations of the moment conditions. The Sargan statistic converges in distribution and therefore is $O_p(1)$.
	
	\paragraph{Proof of Proposition \ref{prop:Sargan-GlobValidLocInvalidK=0.5}}
	With $\bm{\tau} = o_p(1)$, as before
	$$\widehat{\mathbf{u}}(\widehat{\bm{\theta}}_{\mathcal{I}})^{\prime}\widehat{\mathbf{u}}(\widehat{\bm{\theta}}_{\mathcal{I}})/n \rightarrow \sigma_u^2$$
	Following equation \eqref{eq:RewriteEmpiricalMoments-GlobalValidLocalViolations}, the root of the numerator with local invalidity becomes
	\begin{equation}\label{eq:RootOfNumerator-BorderCase-GlobalValidLocalViolations}
		(\mathbf{Z}^{\prime}\mathbf{Z})^{-1/2}\mathbf{Z}^{\prime}	\widehat{\mathbf{u}}(\widehat{\bm{\theta}}_{\mathcal{I}}) = \mathbf{ M}(\mathbf{Z}^{\prime}\mathbf{Z})^{-1/2}\mathbf{Z}^{\prime}(\mathbf{u} + \mathbf{Z}\frac{\mathbf{c}}{\sqrt{n}})
	\end{equation}
	where as before $\mathbf{M}=\mathbf{I} - (\mathbf{Z}^{\prime}\mathbf{Z})^{-1/2}\mathbf{Z}^{\prime}{\mathbf{X}_\mathcal{I}}({\mathbf{X}_\mathcal{I}}^{\prime} \mathbf{P}_Z{\mathbf{X}_\mathcal{I}})^{-1}{\mathbf{X}_\mathcal{I}}^{\prime} \mathbf{Z} (\mathbf{Z}^{\prime}\mathbf{Z})^{-1/2}$ is an idempotent projection matrix that goes to $\mathbf{A}$ in probability. Still,
	\begin{equation*}
		\frac{(\mathbf{Z}^{\prime}\mathbf{Z})^{-1/2}\mathbf{Z}^{\prime}\mathbf{u}}{\sqrt{\widehat{\mathbf{u}}(\widehat{\bm{\theta}}_{\mathcal{I}})^{\prime}\widehat{\mathbf{u}}(\widehat{\bm{\theta}}_{\mathcal{I}})/n}} \overset{d}{\rightarrow} N(0, \mathbf{I})
	\end{equation*}
	and 
	\begin{equation*}
		\frac{(\mathbf{Z}^{\prime}\mathbf{Z})^{-1/2}\mathbf{Z}^{\prime}\mathbf{Z}\frac{\mathbf{c}}{\sqrt{n}}}{\sqrt{\widehat{\mathbf{u}}(\widehat{\bm{\theta}}_{\mathcal{I}})^{\prime}\widehat{\mathbf{u}}(\widehat{\bm{\theta}}_{\mathcal{I}})/n}} = \frac{(\frac{\mathbf{Z}^{\prime}\mathbf{Z}}{n})^{-1/2}\frac{\mathbf{Z}^{\prime}\mathbf{Z}}{n}\mathbf{c}}{\sqrt{\widehat{\mathbf{u}}(\widehat{\bm{\theta}}_{\mathcal{I}})^{\prime}\widehat{\mathbf{u}}(\widehat{\bm{\theta}}_{\mathcal{I}})/n}} \overset{P}{\rightarrow} \frac{1}{\sigma_{u}} \mathbf{Q}_{ZZ}^{1/2}\mathbf{c}
	\end{equation*}
	and hence the second part of equation \eqref{eq:RootOfNumerator-BorderCase-GlobalValidLocalViolations} converges by Slutsky's Theorem:
	$$\frac{(\mathbf{Z}^{\prime}\mathbf{Z})^{-1/2}\mathbf{Z}^{\prime}(\mathbf{u} + \mathbf{Z}\frac{\mathbf{c}}{\sqrt{n}})}{\sqrt{\widehat{\mathbf{u}}(\widehat{\bm{\theta}}_{\mathcal{I}})^{\prime}\widehat{\mathbf{u}}(\widehat{\bm{\theta}}_{\mathcal{I}})/n}} \overset{d}{\rightarrow} N\left(\frac{1}{\sigma_{u}} \mathbf{Q}_{ZZ}^{1/2}\mathbf{c}, \mathbf{I}\right)$$
	By Theorem 2 in \citet[][p. 57-59]{Searle1971Linear}: when $\mathbf{x} \sim N(\bm{\mu}, \mathbf{I})$ and $\mathbf{A}$ is idempotent, $\mathbf{x}^{\prime}\mathbf{Ax} \sim \chi^2_{J-K-1}(\bm{\mu}^{\prime}\mathbf{A}\bm{\mu})$, where $\bm{\mu}^{\prime}\mathbf{A}\bm{\mu}$ is the non-centrality parameter. For the Sargan statistic, this implies:
	\begin{equation*}
		Sar\left(\widehat{\bm{\theta}}_{\mathcal{I}}\right) \overset{d}{\rightarrow} \chi^2_{J - K_{\mathcal{I}} - 1}\left(\frac{1}{\sigma_{u}^2} \mathbf{c}^{\prime}{\mathbf{Q}_{ZZ}^{1/2}}^{\prime}\mathbf{A} \mathbf{Q}_{ZZ}^{1/2}\mathbf{c}\right) \text{\qed}
	\end{equation*}

	\paragraph*{Major violations: $\bm{\kappa < 1/2}$}
	Now, consider the case $\bm{\tau} = \frac{\mathbf{c}}{n^\kappa}$ with $0 < \kappa < 1/2$, i.e.  $\bm{\tau} > \frac{\mathbf{c}}{\sqrt{n}}$. 
	\begin{proposition}\label{prop:Sargan-GlobValidLocInvalidK<1/2}
		If $\bm{\tau} > \frac{\mathbf{c}}{\sqrt{n}}$, then 
		\begin{equation}\label{eq:Sargan-BeyondBorder-GlobalValidLocalViolations}
			\frac{Sar\left(\widehat{\bm{\theta}}_{\mathcal{I}}\right)}{n^{1-2\kappa}} \overset{P}{\rightarrow} \frac{1}{\sigma_{u}^2} \mathbf{c}^{\prime}\mathbf{Q}_{ZZ}^{1/2'}\mathbf{A} \mathbf{Q}_{ZZ}^{1/2}\mathbf{c} \text{}
		\end{equation}
	\end{proposition}
	\noindent Hence as $Sar\left(\widehat{\bm{\theta}}_{\mathcal{I}}\right) = O_p(n^{1-2\kappa})$, for critical values that fulfill $\xi_n = o(n^{\delta})$ with $\delta > 1-2\kappa$ and $\xi_n = o(n)$, the Sargan test still accepts asymptotically.
	
	\paragraph{Proof of Proposition \ref{prop:Sargan-GlobValidLocInvalidK<1/2}}
	As before, since $\bm{\tau} = o_p(1)$:
	
	$$\widehat{\mathbf{u}}(\widehat{\bm{\theta}}_{\mathcal{I}})^{\prime}\widehat{\mathbf{u}}(\widehat{\bm{\theta}}_{\mathcal{I}})/n \overset{P}{\rightarrow} \sigma_u^2 \text{.}$$
	The root of the numerator, ignoring $\mathbf{M}$, becomes 
	
	\begin{equation*}
		\frac{(\mathbf{Z}^{\prime}\mathbf{Z})^{-1/2}\mathbf{Z}^{\prime}(\mathbf{u} + \mathbf{Z}\frac{\mathbf{c}}{n^\kappa})}{\sqrt{\widehat{\mathbf{u}}(\widehat{\bm{\theta}}_{\mathcal{I}})^{\prime}\widehat{\mathbf{u}}(\widehat{\bm{\theta}}_{\mathcal{I}})/n}} 
		= \frac{(\frac{\mathbf{Z}^{\prime}\mathbf{Z}}{n})^{-1/2}(\frac{\mathbf{Z}^{\prime}\mathbf{u}}{\sqrt{n}} + \frac{\mathbf{Z}^{\prime}\mathbf{Z}}{n}\frac{\mathbf{c}}{n^{\kappa-1/2}})}{\sqrt{\widehat{\mathbf{u}}(\widehat{\bm{\theta}}_{\mathcal{I}})^{\prime}\widehat{\mathbf{u}}(\widehat{\bm{\theta}}_{\mathcal{I}})/n}} = n^{1/2 - \kappa}\frac{(\frac{\mathbf{Z}^{\prime}\mathbf{Z}}{n})^{-1/2}(\frac{\mathbf{Z}^{\prime}\mathbf{u}}{n^{1-\kappa}} + \frac{\mathbf{Z}^{\prime}\mathbf{Z}}{n}\mathbf{c})}{\sqrt{\widehat{\mathbf{u}}(\widehat{\bm{\theta}}_{\mathcal{I}})^{\prime}\widehat{\mathbf{u}}(\widehat{\bm{\theta}}_{\mathcal{I}})/n}} 
	\end{equation*}
	\noindent $(\frac{\mathbf{Z}^{\prime}\mathbf{Z}}{n})^{-1/2}(\frac{\mathbf{Z}^{\prime}\mathbf{u}}{n^{1-\kappa}})$ would converge in distribution when $\kappa=1/2$ and goes to zero in probability when $\kappa<1/2$. This follows because it converges to a normally distributed random variable scaled by $n^\nu$ with $\nu>0$. The term $\frac{(\frac{\mathbf{Z}^{\prime}\mathbf{Z}}{n})^{-1/2}(\frac{\mathbf{Z}^{\prime}\mathbf{Z}}{n}\mathbf{c})}{\sqrt{\widehat{\mathbf{u}}(\widehat{\bm{\theta}}_{\mathcal{I}})^{\prime}\widehat{\mathbf{u}}(\widehat{\bm{\theta}}_{\mathcal{I}})/n}} $ converges in probability to $\frac{1}{\sigma_{u}} \mathbf{Q}_{ZZ}^{1/2}\mathbf{c}$, hence
	
	\begin{equation}
		\frac{Sar\left(\widehat{\bm{\theta}}_{\mathcal{I}}\right)}{n^{1-2\kappa}} \overset{P}{\rightarrow} \frac{1}{\sigma_{u}^2} \mathbf{c}^{\prime}\mathbf{Q}_{ZZ}^{1/2'}\mathbf{A} \mathbf{Q}_{ZZ}^{1/2}\mathbf{c} \text{\qed}
	\end{equation}
	
	\subsubsection{Mixture of globally valid and invalid IVs, locally valid}\label{sec:Mixture-NoViolations}
	
	Next, assume that the incorrect set of instruments has been selected as valid. 
	The model where some IVs have been wrongly included as valid is denoted as 
	\begin{equation}\label{eq:OracleModel-Near}
		\mathbf{y} = \mathbf{d}\beta + \mathbf{Z}_\mathcal{A}\bm{\alpha}_\mathcal{A} + \bm{\xi} = \mathbf{X}_\mathcal{A}\bm{\theta}_\mathcal{A} + \bm{\xi}
	\end{equation}
	A mixture of instruments from different groups has been selected instead. For example, some invalid instruments have been selected as valid and some valid IVs have been correctly selected. We can rewrite Equation \eqref{eq:OracleModel-Near}  with $\bm{\xi} = \mathbf{Z}_1\bm{\alpha}_1 + \mathbf{u}$, $\mathbf{Z}_1$ denoting the invalid IVs selected as valid:
	\begin{equation*}
		\mathbf{y} = \mathbf{X}_{\mathcal{A}} \bm{\theta}_{\mathcal{A}} + \mathbf{Z}_1\bm{\alpha}_1 + \mathbf{u}\text{.}
	\end{equation*}
	
	\begin{proposition}\label{prop:Sargan-MixtureLocValid}
		When testing a mixture of globally valid and invalid IVs, if $\bm{\tau} = o_p(1)$, then
		$$Sar\left(\widehat{\bm{\theta}}_{\mathcal{A}}\right) = O_p(n)$$
	\end{proposition}
	
	\paragraph{Proof of Proposition \ref{prop:Sargan-MixtureLocValid}}
	
	\noindent Under model \eqref{eq:OracleModel-Near} the estimator becomes
	\begin{align*}
		\begin{split}
			\widehat{\bm{\theta}}_{\mathcal{A}} & = ({\mathbf{X}_{\mathcal{A}}}^{\prime} \mathbf{P}_Z \mathbf{X}_{\mathcal{A}})^{-1} {\mathbf{X}_{\mathcal{A}}}^{\prime} \mathbf{P}_Z \mathbf{y} = ({\mathbf{X}_{\mathcal{A}}}^{\prime} \mathbf{P}_Z \mathbf{X}_{\mathcal{A}})^{-1} {\mathbf{X}_{\mathcal{A}}}^{\prime} \mathbf{P}_Z (\mathbf{X}_\mathcal{A}\bm{\theta}_\mathcal{A} + \bm{\xi}) \\  
			& = \bm{\theta}_\mathcal{A} + ({\mathbf{X}_{\mathcal{A}}}^{\prime} \mathbf{P}_Z \mathbf{X}_{\mathcal{A}})^{-1} {\mathbf{X}_{\mathcal{A}}}^{\prime} \mathbf{P}_Z \bm{\xi}
		\end{split}
	\end{align*}
	The residual becomes
	\begin{equation*}
		\hat{\bm{\xi}} = \widehat{\bm{\xi}}(\widehat{\bm{\theta}}_{\mathcal{A}}) = \bm{\xi} - \mathbf{X}_{\mathcal{A}}({\mathbf{X}_{\mathcal{A}}}^{\prime} \mathbf{P}_Z \mathbf{X}_{\mathcal{A}})^{-1} {\mathbf{X}_{\mathcal{A}}}^{\prime} \mathbf{P}_Z \bm{\xi}
	\end{equation*}
	\noindent To see to what the inner product of the residual converges, we look at each element of the product: 
	Note that $\bm{\xi}^{\prime}\bm{\xi}/n \overset{P}{\rightarrow} \bm{\alpha}_1' \mathbf{Q}_{Z_1Z_1} \bm{\alpha}_1 + \sigma_u^2$. The terms $\bm{\xi}^{\prime} \mathbf{Z}_\mathcal{A}/n$ and $\bm{\xi}^{\prime} \mathbf{X}_\mathcal{A}/n$ all converge in probability to finite vectors and hence 
	\begin{align}\label{eq:Mixture-Terms}
		\begin{split}
			&\bm{\xi}^{\prime}\mathbf{X}_{\mathcal{A}}({\mathbf{X}_{\mathcal{A}}}^{\prime} \mathbf{P}_Z \mathbf{X}_{\mathcal{A}})^{-1} {\mathbf{X}_{\mathcal{A}}}^{\prime} \mathbf{P}_Z \bm{\xi}/n\\  
			&\text{ and } \\
			&\bm{\xi}^{\prime}\mathbf{P}_Z\mathbf{X}_{\mathcal{A}} ({\mathbf{X}_{\mathcal{A}}}^{\prime} \mathbf{P}_Z \mathbf{X}_{\mathcal{A}})^{-1}{\mathbf{X}_{\mathcal{A}}}^{\prime}\mathbf{X}_{\mathcal{A}}({\mathbf{X}_{\mathcal{A}}}^{\prime} \mathbf{P}_Z \mathbf{X}_{\mathcal{A}})^{-1} {\mathbf{X}_{\mathcal{A}}}^{\prime} \mathbf{P}_Z \bm{\xi}/n
		\end{split}
	\end{align} 
	also do. Denote the first term as $C_1$ and the second term as $C_2$. Let $C=- 2C_1 + C_2$. Then, the denominator of the Sargan-statistic converges in probability to $\sigma_u^2$ plus an inconsistency
	\begin{equation}\label{eq:ErrorVariance-Mixture}
		\widehat{\bm{\xi}}(\widehat{\bm{\theta}}_{\mathcal{A}})'\widehat{\bm{\xi}}(\widehat{\bm{\theta}}_{\mathcal{A}})/n \overset{P}{\rightarrow} \sigma_u^2 + \bm{\alpha}_1'\mathbf{Q}_{Z_1Z_1}\bm{\alpha}_1 + C
	\end{equation}
	The root of the numerator is
	\begin{align*}
		(\mathbf{Z}^{\prime}\mathbf{Z})^{-1/2}\mathbf{Z}^{\prime}\hat{\bm{\xi}} & = (\mathbf{Z}^{\prime}\mathbf{Z})^{-1/2}\mathbf{Z}^{\prime}(\bm{\xi} - \mathbf{X}_\mathcal{A}({\mathbf{X}_\mathcal{A}}^{\prime} \mathbf{P}_Z \mathbf{X}_\mathcal{A})^{-1} {\mathbf{X}_\mathcal{A}}^{\prime} \mathbf{P}_Z \bm{\xi}) \\
		& = (\mathbf{I} - (\mathbf{Z}^{\prime}\mathbf{Z})^{-1/2}\mathbf{Z}^{\prime}\mathbf{X}_\mathcal{A}({\mathbf{X}_\mathcal{A}}^{\prime} \mathbf{P}_Z\mathbf{X}_\mathcal{A})^{-1}{\mathbf{X}_\mathcal{A}}^{\prime} \mathbf{Z} (\mathbf{Z}^{\prime}\mathbf{Z})^{-1/2} )(\mathbf{Z}^{\prime}\mathbf{Z})^{-1/2}\mathbf{Z}^{\prime}\bm{\xi}\\
		& = (\mathbf{I} - (\mathbf{Z}^{\prime}\mathbf{Z})^{-1/2}\mathbf{Z}^{\prime}\mathbf{X}_\mathcal{A}({\mathbf{X}_\mathcal{A}}^{\prime} \mathbf{P}_Z\mathbf{X}_\mathcal{A})^{-1}{\mathbf{X}_\mathcal{A}}^{\prime} \mathbf{Z} (\mathbf{Z}^{\prime}\mathbf{Z})^{-1/2} ) (\mathbf{Z}^{\prime}\mathbf{Z})^{-1/2}\mathbf{Z}^{\prime}(\mathbf{Z}_1\bm{\alpha}_1 + \mathbf{u})\text{.}
	\end{align*}
	For the expression behind the projection matrix it holds that
	\begin{equation*}
		\frac{1}{\sqrt{n}}(\mathbf{Z}^{\prime}\mathbf{Z})^{-1/2}\mathbf{Z}^{\prime}(\mathbf{Z}_1\bm{\alpha}_1 + \mathbf{u}) \overset{P}{\rightarrow} \mathbf{Q}_{ZZ}^{-1/2} \mathbf{Q}_{ZZ_1}\bm{\alpha}_1
	\end{equation*}
	Therefore, after dividing by $n$ (because the last term with $\frac{1}{\sqrt{n}}$ appears twice), all of the matrices in the denominator and numerator of the Sargan test statistic converge in probability:
	\begin{equation}\label{eq:Sargan-Mixture}
		\frac{Sar\left(\widehat{\bm{\theta}}_{\mathcal{A}}\right)}{n} \overset{P}{\rightarrow} \frac{\bm{\alpha}_1' {\mathbf{Q}_{ZZ_1}}^{\prime} {\mathbf{Q}_{ZZ}^{-1/2}}^\prime \mathbf{A} \mathbf{Q}_{ZZ}^{-1/2} \mathbf{Q}_{ZZ_1} \bm{\alpha}_1}{\sigma_u^2 + \bm{\alpha}_1'\mathbf{Q}_{Z_1Z_1}\bm{\alpha}_1 + C}
	\end{equation}
	This means that  $Sar\left(\widehat{\bm{\theta}}_{\mathcal{A}}\right) = O_p(n)$. $\qed$
	
	\subsubsection{Mixture of globally valid and invalid IVs, with local violations}
	
	Next, assume we have selected instruments incorrectly, and hence we have a mixture of valid and invalid instruments, but there are local violations $\tau_j$. The error in the model with too few IVs selected as valid is now 
	\begin{equation}
		\bm{\xi} = \mathbf{Z}_1\bm{\alpha}_1 + \mathbf{Z}\bm{\tau} + \mathbf{u}\text{.}
	\end{equation}
	
	\begin{proposition}\label{prop:Sargan-MixtureLocInvalid}
		When testing a mixture of globally valid and invalid IVs, if $\bm{\tau} = o_p(1)$, then
		$$Sar\left(\widehat{\bm{\theta}}_{\mathcal{A}}\right) = O_p(n)\textbf{.}$$
	\end{proposition}
	
	\paragraph{Proof of Proposition \ref{prop:Sargan-MixtureLocInvalid}}
	The inner product of the error term divided by $n$ is $\bm{\xi}^{\prime}\bm{\xi}/n \overset{P}{\rightarrow} \bm{\alpha}_1^{\prime} \mathbf{Q}_{Z_1Z_1} \bm{\alpha}_1 + \bm{\tau}^{\prime} \mathbf{Q}_{ZZ} \bm{\tau}+ \sigma_u^2$. Note that if $\bm{\tau} = o_p(1)$, this inner product as well as the terms $\bm{\xi}^{\prime} \mathbf{Z}_\mathcal{A}/n$ and $\bm{\xi}^{\prime} \mathbf{X}_\mathcal{A}/n$ and the terms in \eqref{eq:Mixture-Terms} still converge to the same finite matrices and scalars as before. 
	Equation \eqref{eq:ErrorVariance-Mixture} therefore still holds. Also, if $\bm{\tau}$ vanishes, for the root of the numerator it still holds that
	\begin{equation*}
		\frac{1}{\sqrt{n}}(\mathbf{Z}^{\prime}\mathbf{Z})^{-1/2}\mathbf{Z}^{\prime}(\mathbf{Z}_1\bm{\alpha}_1 + \mathbf{Z}\bm{\tau} + \mathbf{u}) \overset{P}{\rightarrow} \mathbf{Q}_{ZZ}^{-1/2} \mathbf{Q}_{ZZ_1}\bm{\alpha}_1 \text{.}
	\end{equation*}
	Hence it again holds that the Sargan test statistic divided by $n$ converges in probability to the same expression as in \eqref{eq:Sargan-Mixture} and the Sargan statistic is $O_p(n)$ $\qed$
	
	\section{Additional Simulations: Single Regressor Case}\label{sec:SimulationsAppendix}
	\subsection{Simulations With Strong Instruments}\label{sec:SimulationsOneRegressor}

\begin{table}[!t] \centering 
  \caption{Simulation Results with One Regressor} 
  \label{tab:frank} 
\begin{tabular}{@{\extracolsep{5pt}} ccccccc} 
\toprule
 & MAE & SD & \# invalid & p allinv & Coverage & p oracle \\ 
\midrule
\multicolumn{7}{c}{\textbf{n=500}}  \\ 
oracle & 0.016 & 0.025 & 12     & 1     & 0.929 & 1 \\ 
naive  & 1.056 & 0.049 & 0      & 0     & 0     & 0 \\ 
HT     & 1.165 & 0.127 & 12.696 & 0     & 0     & 0 \\ 
CIM    & 0.017 & 0.267 & 12.023 & 0.987 & 0.906 & 0.966 \\ 
AHC    & 0.016 & 0.179 & 12.049 & 0.989 & 0.912 & 0.983 \\ 
\midrule
\multicolumn{7}{c}{\textbf{n=1000}}  \\ 
oracle & 0.012 & 0.017 & 12     & 1     & 0.953 & 1 \\ 
naive  & 1.058 & 0.034 & 0      & 0     & 0     & 0 \\ 
HT     & 1.374 & 0.114 & 18.205 & 0     & 0.001 & 0 \\ 
CIM    & 0.012 & 0.017 & 12.015 & 1     & 0.948 & 0.986 \\ 
AHC    & 0.012 & 0.135 & 12.052 & 0.991 & 0.936 & 0.980 \\ 
\midrule
\multicolumn{7}{c}{\textbf{n=2000}}  \\ 
oracle & 0.008 & 0.012 & 12     & 1     & 0.943 & 1 \\ 
naive  & 1.059 & 0.025 & 0      & 0     & 0     & 0 \\ 
HT     & 0.010 & 0.384 & 12.679 & 0.885 & 0.864 & 0.708 \\ 
CIM    & 0.008 & 0.012 & 12.013 & 1     & 0.938 & 0.988 \\ 
AHC    & 0.008 & 0.160 & 12.039  & 0.993& 0.931 & 0.984 \\
\hline \\[-1.8ex] 
\multicolumn{7}{c}{\begin{minipage}{0.75\textwidth} \footnotesize 
                This table reports median absolute error, standard deviation, 
                number of IVs selected as invalid, frequency with which all invalid 
                IVs have been selected as invalid, coverage rate of the 95 \% confidence interval 
                and frequency with which oracle model has been selected. 
                The true coefficient is $\beta=0$. WLHB setting and invalid weaker setting
                are described in the text. 1000 repetitions per setting.
                \end{minipage}} \\ 
\end{tabular} 
\end{table}

	This section provides details on the simulation results for one regressor and strong instruments. The first-stage parameters are given by $\bm{\gamma} = c_\gamma \times  \bm{\iota}_{21} $ and we set $c_\gamma=0.4$. The invalidity vector is as in the main text: $\bm{\alpha} =  \left( \bm{\iota}_6', \, \, \, 0.5 \bm{\iota}_6', \,\,\, \bm{0}_9' \right)'$. 
	The IV selection and estimation results are presented in Table \ref{tab:frank} for sample sizes $n = 500,$ $1000, 2000$ for $1000$ Monte Carlo replications. 
	
	For $n = 500$, the oracle 2SLS estimator (\textit{oracle}), which uses only the valid IVs and controls for the truly invalid ones, has the lowest MAE at $0.016$ and the coverage rate of the 95 \% confidence interval is at $0.929$. The naive 2SLS estimator (\textit{naive}) which treats all candidates instruments as valid irrespective of their validity, however, has a much larger median absolute error of about $1.056$ and its 95 \% confidence interval never covers the true value. This does not change even when increasing the sample size to $2000$, as expected. When using the HT method with 500 observations, the MAE is even larger than that of the naive 2SLS estimator and the method never chooses the oracle model, leading none of the confidence intervals to cover the true value. This is in line with the IV selection results: the frequency of including all invalid instruments as invalid, and that of selecting the oracle model are $0$. When using CIM, the MAE is already low when $n=500$, the number of IVs chosen as invalid is close to 12, the frequency with which the oracle model is selected is at $0.966$, and the coverage rate is $0.906$. Results are very similar for our AHC method. When increasing the sample size, the performance improves for all three selection methods. For CIM and AHC, the MAE is equal to that of the oracle estimator both for $n = 1000$ and $n = 2000$, and the probabilities to select the oracle model are close to one, while for HT it is lower, showing that CIM and AHC have better performance in this setting when the sample size is relatively small.
	
	\subsection{Some Weak Instruments Among Candidates}\label{sec:SimulationsP1Weak}
	We examine the performance of the AHC method when there are weak instruments among the candidates and there is a single regressor. We compare performance with the HT and CIM methods.
	
	Firstly, consider the same simulation setting as in Section \ref{sec:SimulationsOneRegressor} but with the following variations:
	\begin{itemize}
		\item Design 1: All the 12 invalid instruments are weak, and all the 9 valid instruments are strong: $\bm{\gamma} = c_{\gamma}\left(\bm{\iota}_{12}'C/\sqrt{n}, \,\,\,\bm{\iota}_9'\right)'$.
		\item Design 2: All the 12 invalid instruments are weak, and 4 out of 9 of the valid instruments are weak: $\bm{\gamma} = c_{\gamma}\left(\bm{\iota}_{16}'C/\sqrt{n}, \,\,\,\bm{\iota}_5'\right)'$.
		\item Design 3: Both the valid and invalid IVs are mixtures of weak and strong IVs.
		\begin{itemize}
			\item a). Strong and valid instruments still form the largest group:\\ $\bm{\gamma} = c_{\gamma}\left(\bm{\iota}_6', \,\,\, \bm{\iota}_{7}'C/\sqrt{n}, \,\,\,\bm{\iota}_8'\right)'$.
			\item b). Strong and valid instruments do not form the (strictly) largest group:\\ $\bm{\gamma} = c_{\gamma}\left(\bm{\iota}_6', \,\,\, \bm{\iota}_{9}'C/\sqrt{n}, \,\,\,\bm{\iota}_6'\right)'$.
		\end{itemize}
	\end{itemize}
	All the other parameters are the same as in Section \ref{sec:SimulationsOneRegressor}. We 
	fix the sample size to $n = 2000$ and present the results in Table \ref{tab:weaksingle}.
	
\begin{table}[!htbp] \centering 
  \caption{Some Weak Instruments with One Regressor} 
  \label{tab:weaksingle} 
\begin{tabular}{@{\extracolsep{5pt}} ccccccc} 
\toprule
 & MAE & \# invalid &    p allinv & strongvalid & weakin & weakva \\ 
\hline \\[-1.8ex] 
\multicolumn{7}{c}{\textbf{Design 1}} \\ 
oracle   & 0.008  & 12     & 1     & 1     & 1     &-  \\ 
HT       & 0.008  & 12.000 & 1     & 1     & 1     &-  \\ 
CIM      & 35.112 & 13.289 & 0.024 & 0     & 0.024 &- \\ 
AHC      & 0.008  & 12.028 & 1     & 0.988 & 1     &-  \\ 
\midrule
\multicolumn{7}{c}{\textbf{Design 2}} \\
oracle   & 0.013  & 16     & 1     & 1     & 1     &1  \\ 
HT       & 0.013  & 15.951 & 1     & 1     & 1     &0.952  \\ 
CIM      & 33.646 & 12.806 & 0.027 & 0     & 0.027 &0.527 \\ 
AHC      & 0.012  & 12.445 & 0.999 & 0.997 & 0.999 &0.002 \\ 
\midrule
\multicolumn{7}{c}{\textbf{Design 3a}} \\
oracle   & 0.008  & 13     & 1     & 1     & 1     &1 \\ 
HT       & 0.008  & 13.164 & 1     & 0.842 & 1     &0.984  \\ 
CIM      & 14.497 & 16.772 & 0.351 & 0.002 & 0.467 &0.691\\ 
AHC      & 0.008  & 12.323 & 0.998 & 0.992 & 1     &0.306  \\ 
\midrule
\multicolumn{7}{c}{\textbf{Design 3b}} \\
oracle   & 0.011  & 15     & 1     & 1     & 1     &1  \\ 
HT       & 0.929  & 10.511 & 0.053 & 0.870 & 0.999 &0.961  \\ 
CIM      & 13.636 & 16.500 & 0.277 & 0.008 & 0.462 &0.421 \\ 
AHC      & 0.013  & 12.766 & 0.847 & 0.847 & 1     &0.002  \\ 
\bottomrule \\[-1.8ex] 
\multicolumn{7}{c}{\begin{minipage}{0.78\textwidth} \footnotesize 
                This table reports median absolute error, 
                number of IVs selected as invalid, frequency of all invalid 
                IVs selected as invalid, frequency of all valid and strong instruments selected as valid, frequency of all weak invalid instruments selected as invalid, and frequency of all weak valid instruments as invalid. 1000 repetitions per setting.
                \end{minipage}} \\ 
\end{tabular} 
\end{table}
	
	\noindent From the simulation results we can see that with weak instruments, the CI method can be problematic: the frequencies of selecting all invalid instruments as invalid (\textit{p allinv}) are low in all settings (lowest at $0.024$ in Design 1 and Design 2, and highest at $0.351$ in Design 3a), meaning that it almost always includes invalid instruments as valid. Consequently, the MAE for CIM is very large (and much larger than those of the oracle, HT and AHC). Particularly, CIM tends to select weak invalid instruments as valid (low \textit{weakin} in all settings). This is because their confidence intervals are wide. Thus, most of them will be overlapping with all other confidence intervals and the resulting largest group of mutually overlapping confidence intervals (the selected set of valid IVs) tends to always contain the weak invalid IVs.
	
	The HT method performs well in all designs where the plurality rule holds (Design 1 to Design 3a). It selects all weak instruments (both valid and invalid) as invalid with probability almost equal to 1. Also, it has high frequencies of selecting all strong and valid instruments as valid. It can be seen that if the strong and valid instruments form the largest group, the voting mechanism of HT can select the oracle model. The good IV selection performance of HT in presence of weak instruments is mainly due to its first-stage IV selection, the pre-screening of weak IVs (either valid or invalid). In line with the selection performance, the MAE of HT is very close to that of the oracle model.
	In Design 3b, however, the plurality rule does not hold anymore: there is a tie between the group of strong and valid instruments, and strong and invalid instruments. In this situation, the voting mechanism does not perform well as \textit{p allinv} is only at $0.053$. This results in a significantly larger MAE than the oracle model.
	
	The AHC performs well in general and has similar MAE to the oracle model in all settings. For Design 1, 2 and 3a, it guarantees that all the invalid IVs are selected as invalid with \textit{p allinv} and \textit{weakin} close to 1. In terms of valid IVs, all the strong valid instruments are included as valid with high frequencies (\textit{strongvalid} close to 1). However, for weak valid instruments, some of them are selected as valid with relatively low \textit{weakva}. This is because the just-identified estimators of the weak valid instruments may not be too far away from those of the strong and valid instruments, thus in some cases they are not totally separated by the algorithm. But this is not the major concern, as for weak valid instruments, the algorithm would only keep those whose estimators are not severely distorted, and the effect of the selected weak instruments on the resulting post-selection IV estimator is limited. This can be seen from the MAE of AHC which are very close to those of the oracle models even with low \textit{weakva}. It should be noted that in Design 3b where there are two largest groups, AHC outperforms HT with a frequency of $0.847$ of including all invalid IVs as invalid. Alternatively, AHC can report both groups. 
	Overall, we find that AHC clearly outperforms CIM in all settings with weak IVs and it also outperforms HT in the case where plurality is not strictly fulfilled. The main drawback of AHC is that it tends to include weak valid IVs as valid, which can be potentially complemented by the weak IV pre-screening procedure in HT.
	
	\subsection{Local Violations}
	
	\begin{table}[!htbp] \centering 
	\caption{Local Violations with One Regressor} 
	\label{tab:localviolation} 
	\begin{tabular}{@{\extracolsep{5pt}} p{1.2cm}p{1cm}p{1cm}p{1.7cm}p{1.5cm}p{1.7cm}p{1.5cm}p{1cm}} 
		\toprule
		& MAE &SD & \# invalid &    p allinv & global viol &p allins   & coverage \\ 
		\hline \\[-1.8ex] 
\multicolumn{8}{c}{\textbf{Design 1:}, $\bm{\kappa=3/4}$} \\ 
		oracle   & 0.008  &0.012 & 12     & 1     & 6     & 1       &0.952\\ 
		naive   & 0.383  & 0.014& 0     & 0     & 0     & 0       &0.000\\ 
		HT       &0.006   &0.009 &6  & 0     & 6     & 1     &0.943\\ 
		CIM      &0.006  &0.009 &6.016  & 0 & 6     &1   &0.942\\ 
		AHC      & 0.007  &0.011 &6.066  &  0    &6  & 1   &0.932  \\ 
		\midrule
\multicolumn{8}{c}{\textbf{Design 2:}, $\bm{\kappa=1/2}$} \\ 
		oracle   & 0.008 &0.012  & 12     & 1     & 6     & 1     &0.952 \\ 
		naive & 0.395  &0.014 & 0     & 0     & 0     & 0       &0.000\\ 
		HT       & 0.021 &0.009 & 6.001 & 0      & 6     & 1   & 0.410\\ 
		CIM      &0.020 &0.012 &6.256  & 0  &6  &1    &0.448\\ 
		AHC      &0.022  &0.026&6.982  &0  &6  &1   &0.390\\ 
		\midrule
\multicolumn{8}{c}{\textbf{Design 3:}, $\bm{\kappa=1/4}$} \\ 
		oracle   & 0.008 &0.012  & 12     & 1     & 6     & 1     &0.952 \\ 
		naive & 0.482  &0.013 & 0     & 0     & 0     & 0       &0.000\\ 
		HT       & 0.101 &0.047 & 13.156 & 0.039      & 6     & 1   & 0.045\\ 
		CIM      &0.009 &0.020 &11.687  & 0.659  &6  &1    &0.919\\ 
		AHC      &0.009  &0.136&12.124  &0.927  &5.935  &0.989   &0.905\\ 
		\bottomrule \\[-1.8ex] 
		\multicolumn{8}{c}{\begin{minipage}{0.95\textwidth} \footnotesize 
				This table reports median absolute error, standard deviation, 
				number of IVs selected as invalid, frequency of all invalid 
				IVs selected as invalid, number of invalid IVs with global violations selected as invalid, frequency of selecting all the invalid IVs with global violations as invalid, and coverage rate at 5\% significance level. 1000 repetitions per setting.
		\end{minipage}} \\ 
	\end{tabular} 
\end{table}
	
	\begin{table}[!ht] \centering 
	\caption{Post-selection inference using the search and sampling method} 
	\label{tab:SearchAndSampling} 
	\begin{tabular}{@{\extracolsep{5pt}} p{1.2cm}p{1cm}p{1cm}p{1.7cm}p{1.5cm}p{1.7cm}p{1.5cm}p{1cm}} 
		\toprule
		& SD & cover & cover-se &    cover-sa & width & wid-se   & wid-sa \\ 
		\hline \\[-1.8ex] 
\multicolumn{8}{c}{$\bm{\kappa = 1/2}$, $\bm{n = 500}$} \\ 
		oracle   & 0.024   &0.949  & -     & -        & 0.094     & -       &-\\ 
		naive    & 0.027   & 0.000 & -     & -        & 0.099     & -       &-\\ 
		HT       &0.019    &0.364  &1.000  & 1.000    & 0.074     &0.918    &0.389\\ 
		CIM      &0.024    &0.409  &1.000  & 1.000    & 0.076     &0.909    &0.383\\ 
		AHC      & 0.049   &0.353   &1.000  & 0.985   & 0.083     &0.906     &0.379  \\ 
		\midrule
\multicolumn{8}{c}{$\bm{\kappa = 1/2}$, $\bm{n = 1000}$} \\ 
		oracle   & 0.017   &0.951  & -     & -        & 0.066     & -       &-\\ 
		naive    & 0.018   & 0.000 & -     & -        & 0.070     & -       &-\\ 
		HT       &0.013    &0.386  &1.000  & 0.999    & 0.052     &0.608    &0.268\\ 
		CIM      &0.017    &0.407  &1.000  & 1.000    & 0.053     &0.604    &0.265\\ 
		AHC      &0.033    &0.358   &1.000  & 0.989   & 0.057     &0.601    &0.263  \\ 
		\midrule
  \multicolumn{8}{c}{$\bm{\kappa = 1/2}$, $\bm{n = 2000}$} \\ 
		oracle   & 0.012   &0.952  & -     & -        & 0.047     & -       &-\\ 
		naive    & 0.014   & 0.000 & -     & -        & 0.050     & -       &-\\ 
		HT       &0.009    &0.412  &1.000  & 0.999    & 0.037     &0.419    &0.189\\ 
		CIM      &0.012    &0.448  &1.000  & 1.000    & 0.038     &0.415    &0.186\\ 
		AHC      &0.026    &0.390   &1.000  & 0.994   &0.041     &0.413    &0.185  \\ 
		\bottomrule \\[-1.8ex] 
		\multicolumn{8}{c}{\begin{minipage}{0.9\textwidth} \footnotesize 
				This table reports standard deviation (\textit{SD}), 
				coverage rate of the post-selection 2SLS 95\% confidence interval (\textit{cover}), coverage rate of the 95\% confidence interval using the search method (\textit{cover-se}), coverage rate of the 95\% confidence interval using the sampling method (\textit{cover-sa}), width of the post-selection 2SLS 95\% confidence interval, (\textit{width}), width of the 95\% confidence interval using the search method (\textit{wid-se}), width of the 95\% confidence interval using the sampling method (\textit{wid-sa}). 1000 repetitions per setting.
		\end{minipage}} \\ 
	\end{tabular} 
\end{table}	
	
	Consider the same setting as in Appendix \ref{sec:SimulationsOneRegressor} with one endogenous variable, but with the following variations:
	\begin{itemize}
		\item Design 1, minor violations with strong IVs: 6 IVs are globally invalid, while 6 are locally invalid with $\bm{\alpha} = \left( \bm{0}_6', \,\,\bm{\iota}_60.5', \,\, \bm{0}_9'\right)'$, $\bm{\tau} = c_{\alpha}\left(\bm{\iota}_{6}'c/n^{k}, \,\, \bm{0}_{15}'\right)'$ with $\kappa = 3/4, c_\alpha = 1$.
		\item Design 2, mild violations with strong instruments: same as Design 1, but with $\kappa=1/2$
		\item Design 3, strong violations with strong IVs: same as Design 1, but with $\kappa=1/4$
	\end{itemize}
	All the other parameters are the same as in Section \ref{sec:SimulationsOneRegressor}. Here we report two new statistics that report how well a method does at detecting global violations: the number of invalid instruments with strong violations that are selected as invalid (\textit{global violation}) and the frequency of selecting all the invalid instruments with global violations as invalid (\textit{p allins}). The simulation results are presented in Table \ref{tab:localviolation}.
	
	In Design 1, all methods work well in selecting the six IVs with global violations. The naive 2SLS is biased in this setting but the selection methods are not, because keeping IVs with minor violations does not hurt the asymptotic performance of estimators. In Design 2, all three methods perform well in terms of finding global violations, with frequencies of selecting all the IVs with global violations as invalid equal to 1, and the numbers of such IVs selected as invalid equal to 6. However, invalid instruments with local violations, tend to be selected as valid by all three methods, leading to bias and resulting in deviations from the oracle results in terms of MAE. This is what was expected from our results on the Sargan test. 
	In Design 3, global as well as strong local violations are detected by CIM and AHC, with HT having a larger bias and low coverage (0.045). 
	
	In table \ref{tab:SearchAndSampling} we have applied the search and sampling method of \citet{Guo2023Causal} for the case with mild violations. This approach is discussed in section \ref{sec:simlocalviolations}. For cases where the methods in fact make selection mistakes, as modelled by the mild violations, the standard CIs are narrow and coverage is much lower than the nominal level of 95 percent. Once we use the searching and sampling approaches we get much wider CIs which are now even conservative with coverage close to 1. This indicates there might be scope for more work improving post-selection inference. 
	
	\section{Additional Simulations: Multiple Regressors}
	
	\begin{table}[!htbp] \centering 
	\caption{Local Violations with Two Regressors and Weak IVs} 
	\label{tab:localviolationmultiweak} 
	\begin{tabular}{@{\extracolsep{5pt}} p{1.5cm}p{1cm}p{1cm}p{1.7cm}p{1.5cm}p{1.7cm}p{1.5cm}p{0.5cm}} 
		\toprule
		& MAE &SD & \# invalid &    p allinv & global viol &p allins   & coverage \\ 
		\hline \\[-1.8ex] 
\multicolumn{8}{c}{$\bm{\kappa=3/4}$} \\ 
		oracle   & 0.022  &0.032 & 12     & 1     & 6     & 1       &0.932\\ 
		or global   & 0.016  &0.023 & 8     & 2/3     & 6     & 1       &0.934\\ 
		naive   & 0.423  & 0.074& 0     & 0     & 0     & 0       &0.000\\ 
		AHC      & 0.016 &0.026 &7.006  &  0.000    &6.000  & 1.000   &0.922  \\ 
		\midrule
\multicolumn{8}{c}{$\bm{\kappa=1/2}$} \\
		oracle   & 0.022 &0.032  & 12     & 1     & 6     & 1     &0.932 \\ 
		or global   & 0.026  &0.024 & 8     & 2/3     & 6     & 1       &0.752\\ 
		naive & 0.455  &0.077 & 0     & 0     & 0     & 0       &0.000\\ 
	        AHC      &0.034  &0.050&9.382  &0.000  &6.000  &1.000   &0.638\\ 
		\midrule
\multicolumn{8}{c}{$\bm{\kappa=1/4}$} \\
		oracle   & 0.022 &0.032  & 12     & 1     & 6     & 1     &0.932 \\ 
		or global   & 0.216  &0.042 & 8     & 2/3     & 6     & 1       &0.000\\ 
		naive & 0.733  &0.108 & 0     & 0     & 0     & 0       &0.000\\ 
		AHC      &0.024  &0.124&12.868  &0.988  &6.000  &1.000   &0.908\\ 
		\bottomrule \\[-1.8ex] 
		\multicolumn{8}{c}{\begin{minipage}{0.95\textwidth} \footnotesize 
				This table reports median absolute error, standard deviation, 
				number of IVs selected as invalid, frequency of all invalid 
				IVs selected as invalid, number of invalid IVs with global violations selected as invalid, frequency of selecting all the invalid IVs with global violations as invalid, and coverage rate at 5\% significance level. 1000 repetitions per setting.
		\end{minipage}} \\ 
	\end{tabular} 
\end{table}
	
	\noindent As a final simulation exercise, we investigate a setting which combines multiple regressors, local violations and some weak IVs. The setting is the same as in the main text, as in Table \ref{tab:localviolationmulti}, with the same $\bm{\alpha}$ and $\bm{\tau}$, but now we introduce weak IVs. The first two IVs of the globally valid and locally invalid and the first two IVs of the globally invalid and locally invalid are set to be weak IVs: $\gamma_{jp} = C/\sqrt{n}$ for $j \in \{1,2,7,10\}$ and $p \in \{1,2\}$. Once again, Designs 1-3 are defined by $\kappa\in\{3/4, 1/2, 1/4\}$. 
	
	Results are displayed in Table \ref{tab:localviolationmultiweak}. For minor violations in Design 1, all global violations are correctly identified because \textit{global viol} is 6 and \textit{p allins} is 1. The number selected as invalid is 7 instead of 12 and \textit{p allinv} is 0, meaning that locally invalid IVs are not selected as invalid, but bias, SD and coverage are still close to the oracle. In Design 2, with mild violations again some locally invalid IVs are not detected, increasing bias and decreasing coverage to 0.638. As compared to the setting without weak IVs, more locally invalid IVs seem to have been detected, because for the weak IVs bias of the just-identified estimators is exacerbated. This is reflected in a higher coverage in the case with weak IVs. In Design 3 with strong violations, similarly to the case without weak IVs, strong local violations are correctly detected and performance is close to the \textit{oracle} estimator.

\end{appendices}

\end{document}